\documentclass[journal=jpclcd,manuscript=article,layout=onecolumn]{achemso}
\setkeys{acs}{articletitle = true}
\newcommand{\onlinecite}[1]{\citenum{#1}}

\usepackage[version=3]{mhchem} 
\usepackage{graphicx}
\usepackage{upgreek}
\usepackage{amsmath}
\usepackage{amssymb}
\usepackage{chemformula}
\usepackage{xcolor}
\usepackage{nicefrac}
\usepackage{booktabs}
\usepackage{comment}
\usepackage{mwe}
\usepackage{array}
\usepackage{syntonly}
\usepackage[utf8]{inputenc}
\usepackage{enumitem}

\usepackage{caption}
\usepackage{float}
\usepackage{subcaption}
\usepackage{graphicx}
\newcolumntype{m}{>{$} c <{$}}
\usepackage{color}

\def\dd{\mathrm{d}}

\def\beq{\begin{equation}}
\def\eeq{\end{equation}}

\def\cc{\mathrm{c}}

\makeatother

\newcommand\VS[2]{\textcolor{red}{{#2}}}

\author{Timothy J. Daas}
\affiliation{Department of Chemistry \& Pharmaceutical Sciences and Amsterdam Institute of Molecular and Life Sciences (AIMMS), Faculty of Science, Vrije Universiteit, De Boelelaan 1083, 1081HV Amsterdam, The Netherlands}
\author{Eduardo Fabiano}
\affiliation{Institute for Microelectronics and Microsystems (CNR-IMM), Via Monteroni, Campus Unisalento, 73100 Lecce, Italy}
\alsoaffiliation{Center for Biomolecular Nanotechnologies, Istituto Italiano di Tecnologia, Via Barsanti 14, 73010 Arnesano (LE), Italy}
\author{Fabio Della Sala}
\affiliation{Institute for Microelectronics and Microsystems (CNR-IMM), Via Monteroni, Campus Unisalento, 73100 Lecce, Italy}
\alsoaffiliation{Center for Biomolecular Nanotechnologies, Istituto Italiano di Tecnologia, Via Barsanti 14, 73010 Arnesano (LE), Italy}
\author{Paola Gori-Giorgi}
\affiliation{Department of Chemistry \& Pharmaceutical Sciences and Amsterdam Institute of Molecular and Life Sciences (AIMMS), Faculty of Science, Vrije Universiteit, De Boelelaan 1083, 1081HV Amsterdam, The Netherlands}
\author{Stefan Vuckovic}
\affiliation{Physical and Theoretical Chemistry, University of Saarland, 66123 Saarbrücken, Germany}
\alsoaffiliation{Department of Chemistry, University of California, Irvine, CA 92697, USA}
\alsoaffiliation{Department of Chemistry \& Pharmaceutical Sciences and Amsterdam Institute of Molecular and Life Sciences (AIMMS), Faculty of Science, Vrije Universiteit, De Boelelaan 1083, 1081HV Amsterdam, The Netherlands}
\email{svuckovi@uci.edu}

\title{Noncovalent interactions from models for the Møller-Plesset adiabatic connection}

\begin{document}     

\begin{tocentry}
\begin{center}
\includegraphics[width=\columnwidth]{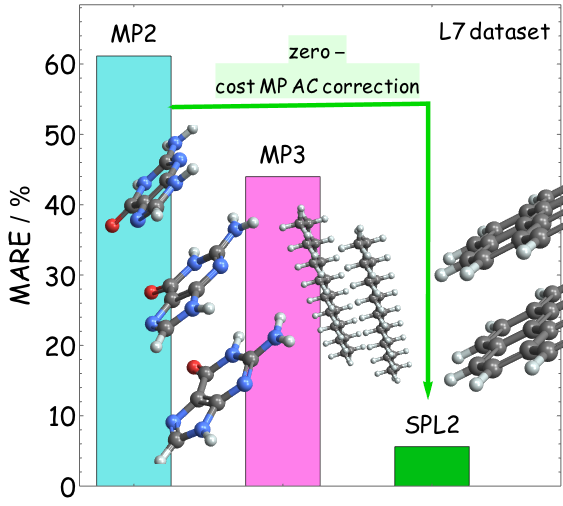}
\end{center}
\end{tocentry}

\begin{abstract}
Given the omnipresence of non-covalent interactions (NCIs), their accurate simulations are of crucial importance across various scientific disciplines. Here we construct accurate models for the description of NCIs by an interpolation along the  Møller-Plesset adiabatic connection (MP AC). Our interpolation approximates the correlation energy, by recovering MP2 at small coupling strengths and the correct large-coupling strength expansion of the MP AC, recently shown to be a functional of the Hartree-Fock density. Our models are size consistent for fragments with non-degenerate ground states, have the same cost as double hybrids and require no dispersion corrections to capture NCIs accurately. These interpolations  greatly reduce large MP2 errors for typical $\pi$-stacking complexes (e.g., benzene-pyridine dimers) and for the L7 dataset. They are also competitive with state-of-the-art dispersion enhanced functionals and can even significantly outperform them for a variety of datasets, such as CT7 and L7.

\end{abstract}

\maketitle



An accurate description of noncovalent interactions (NCIs) is crucial for fields ranging from chemistry, biology and materials science, with a plethora of methods being constantly developed, tested and improved. \cite{hohenstein12,riley11,lao15,sedlak13,grafova10,alhamdani19,grimme16,dubecky16,christensen16,burns11,dilabio13,GAEK10,hobza11,laricchia12,grabowski13,fabiano14,fabiano15,smiga17,fabiano17} Second-order  Møller-Plesset (MP2) perturbation theory has been often considered a relatively safe choice for the treatment of NCIs in chemistry, given its favorable scaling relative to more sophisticated wave-function methods and encouraging early successes in capturing NCIs in small systems \cite{riley12,hobza88}. The described failures of MP2 when applied to NCIs, such as those in stacking complexes, have been often considered accidental. Very recently, Furche and co-workers\cite{Ngu-Fur-JCTC-2020} have shown that MP2 relative errors for NCIs can grow systematically with molecular size, and that the whole MP series may be even qualitatively unsuitable for the description of large non-covalent complexes. On the other hand, double hybrid (DH) functionals, that mix density-functional theory (DFT) semilocal ingredients with a fraction of Hartree-Fock (HF) exchange and MP2 correlation energy, typically worsen the performance of MP2 for NCIs \cite{VucGorDelFab-JPCL-18,Gri-JCP-2006,SchGri-RSC-2007}, unless dispersion corrections are added on top of them. \cite{GAEK10,CalEhlHanNeuSpiBanGri-JCP-2019,CalMewEhlGri-PCCP-2020} A few notable exceptions to this are the XYG$n$ family of functionals~\cite{XYG3,XYG7} and recent DHs developed in Martin's group,\cite{SSM19} which give accuracy improvements over MP2 without requiring additional dispersion corrections. 

\begin{figure}
\includegraphics[width=1\columnwidth]{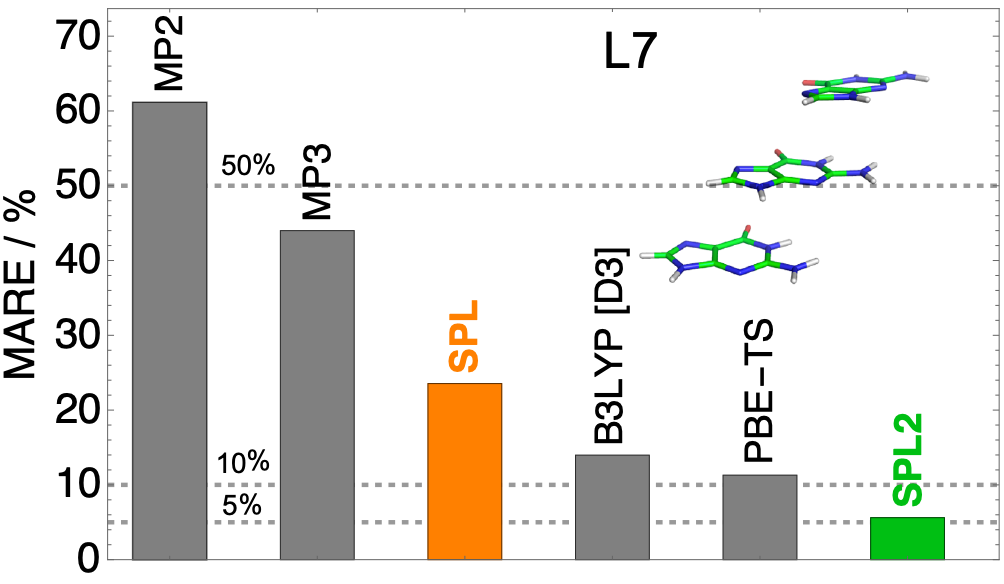}
\caption{The mean absolute relative errors (MARE) for selected methods, where TS stands for the Tkatchenko-Shefler method\cite{TkaSch-PRL-2009}, for the L7 dataset using the reference data of Grimme and co-workers\cite{GriBraBanHan-JCP-2015}.} 
\label{fig:0}
\end{figure}

In this work, we use an adiabatic connection (AC) formalism in which the MP series arises in the weak-coupling limit (hereinafter MP AC) to construct a correction to the MP2 interaction energies, which is guaranteed to be size consistent for the case in which the fragments have a non-degenerate ground-state. We construct this correction using two different strategies both based on an interpolation between MP2 and the strong-coupling limit of the MP AC, which has been recently studied in detail,\cite{SeiGiaVucFabGor-JCP-18,DaaGroVucMusKooSeiGieGor-JCP-20} and shown to be given by functionals of the Hartree-Fock (HF) density with a clear physical meaning.
The resulting method gives major improvements over MP2, despite coming at a negligible extra computational cost. In Fig.~\ref{fig:0} we show, in particular, that a specific interpolation form named `SPL2' -- an extension of the approach of  Seidl, Perdew and Levy\cite{SeiPerLev-PRA-99} (SPL) -- is competitive with state-of-the art electronic structure methods when applied to the challenging L7 dataset.
\cite{SedJanPitRezPulHob-ACS-2013,HamNagBarKalBraTka-unknown-2020,GriBraBanHan-JCP-2015,Ngu-Fur-JCTC-2020}
In what follows, we will describe the theoretical basis for the construction of this new class of functionals based on the MP AC interpolation. We also introduce and analyse the different interpolation schemes within this framework and discuss possible routes to further refinement.

{\it Theory-}
 For the construction of our approximations, we use the Møller-Plesset adiabatic connection (MP AC) framework, whose hamiltonian reads as: 
\begin{equation}\label{eq:HlambdaHF}
	\hat{H}_{\lambda}=\hat{T}+\hat{V}_{\rm ext}+\lambda \hat{V}_{ee}+
	\Big(1-\lambda \Big)
	\Big(\hat{J}+\hat{K} \Big),
\end{equation}
with $\hat{T}$ the kinetic energy, $\hat{V}_{ee}$ the electron-electron repulsion operators. Here $\hat{J}=\hat{J}[\rho^{\rm HF}]$ and $\hat{K}=\hat{K}[\{\phi_i^{\rm HF}\}]$ are the standard Hartree-Fock (HF) Coulomb and exchange operators in terms of the HF density $\rho^{\rm HF}$ and the occupied orbitals $\phi_i^{\rm HF}$, respectively. We denote with $\Psi_\lambda$ the ground-state of $\hat{H}_{\lambda}$, which at $\lambda=1$ corresponds to the physical system and at $\lambda=0$ to the HF Slater determinant. 
In terms of these quantities the traditional quantum-chemical definition of the correlation energy is given by
\beq \label{eq:0h}
E_{\rm c}=\langle  \Psi| \hat{H} | \Psi  \rangle - \langle  \Psi_0 | \hat{H} | \Psi_0 \rangle,
\eeq
with  $\hat{H}=\hat{H}_{\lambda=1}$ and $\Psi=\Psi_{\lambda=1}$  . Applying the Hellman–Feynman
theorem to Eq.~\eqref{eq:HlambdaHF}, one obtains the AC expression for the correlation energy, which reads as\cite{Ngu-Fur-JCTC-2020,Per-IJQC-18,MarBurLoo-JPCM-2021,SeiGiaVucFabGor-JCP-18,DaaGroVucMusKooSeiGieGor-JCP-20}
\beq \label{eq:ac}
E_{\rm c}= \int_0^1 W_{\rm c,\lambda} \dd \lambda,
\eeq
where $W_{\rm c,\lambda}$ is the AC integrand,
\beq \label{eq:wchf}
W_{\cc, \lambda}=\langle \Psi_\lambda
|\hat{V}_{\rm ee} - \hat{J}-\hat{K}|
\Psi_\lambda \rangle 
- \langle  \Psi_0
 |\hat{V}_{\rm ee} - \hat{J}-\hat{K}
 |  \Psi_0\rangle.
\eeq
The small $\lambda$ expansion of $W_{c,\lambda}$ returns the Møller-Plesset (MP) series:
\begin{equation}\label{eq:WHFMP}
    W_{c,\lambda\rightarrow 0}=\sum_{n=2}^\infty n\,E^{{\rm MP}n}_{c}\,\lambda^{n-1},
\end{equation}
where $E^{{\rm MP}n}_{c}$ is the $n$-th order correlation energy from the MP perturbation theory~\cite{Moller1934,SeiGiaVucFabGor-JCP-18}. Very recently, the large-$\lambda$ expansion of $W_{c,\lambda}$ has been shown to have the following form:\cite{DaaGroVucMusKooSeiGieGor-JCP-20}
\begin{equation}  \label{eq:large}
 W_{c,\lambda\rightarrow\infty} = W_{c,\infty} + \frac{W_{\frac{1}{2}}}{\sqrt{\lambda}}+\frac{W_{\frac{3}{4}}}{\lambda^{\frac{3}{4}}}+\dots,
\end{equation}
which is analogous, although with important differences, to the one appearing in the density fixed DFT adiabatic connection (DFT AC).\cite{SeiGorSav-PRA-07,GorVigSei-JCTC-09,SeiGiaVucFabGor-JCP-18} In the DFT AC the correlation energy is given by\cite{LanPer-SSC-75,GunLun-PRB-76}
\begin{align} \label{eq:acdft}
E_{\rm c}^{\rm DFT} [ \rho ] &= \langle  \Psi| \hat{H} | \Psi  \rangle - \langle  \Psi_0^{\rm DFT} | \hat{H} | \Psi_0 ^{\rm DFT}\rangle      \\ \nonumber
   & =            \int_0^1 W_{\rm c,\lambda}^{\rm DFT} \dd \lambda,
\end{align}
where  $\Psi_\lambda^{\rm DFT}$  integrates to the physical density $\rho$ and minimizes the sum of $\hat{T}+\lambda\,\hat{V}_{ee}$. Defined this way, 
$\Psi_\lambda^{\rm DFT}$ is in general equal to $\Psi_\lambda$ only at $\lambda=1$. Furthermore, the density of $\Psi_\lambda$ varies with $\lambda$, whereas the density of  $\Psi_\lambda^{\rm DFT}$ is always the same by construction. The DFT AC integrand $W_{\rm c,\lambda}^{\rm DFT}$ reads as:
\begin{equation} \label{eq:dftint}
W_{\rm c,\lambda}^{\rm DFT}=\langle \Psi_{\lambda}^{\rm DFT}
|\hat{V}_{\rm ee} |
\Psi_\lambda^{\rm DFT} \rangle 
- \langle  \Psi_0^{\rm DFT}
 |\hat{V}_{\rm ee} 
 |  \Psi_0^{\rm DFT} \rangle.
\end{equation}
In the following relations we will also make use of $W_{\lambda}^{\rm DFT}$, defined to include the exchange energy: $W_{\lambda}^{\rm DFT}= W_{\rm c,\lambda}^{\rm DFT} + E_{x}$. The functional $W_{\rm c,\lambda}^{\rm DFT}$ has large~\cite{GorSeiSav-PCCP-08,GorSeiVig-PRL-09,GroKooGieSeiCohMorGor-JCTC-17} and small $\lambda$ \cite{GorLev-PRB-93,GorLev-PRA-94} expansions analogous (but not identical) to those given by Eqs.~\eqref{eq:WHFMP} and~\eqref{eq:large}. 
The large-$\lambda$ limits of the two integrands are related by:\cite{SeiGiaVucFabGor-JCP-18}
\begin{equation} \label{eq:hfleq}
W_{c,\infty} [\rho^{\rm HF}] = W^{\rm DFT}_{\infty}  [\rho^{\rm HF}] + \beta [\rho^{\rm HF}] E_{x}  [\{\phi_i^{\rm HF}\}],
\end{equation}
where dimensionless $\beta [\rho^{\rm HF}]$ is system-dependent, known to satisfy\cite{SeiGiaVucFabGor-JCP-18}
$\beta [\rho^{\rm HF}] \geq 1$,
and for the uniform electron gas (UEG) it is exactly equal to $1$.\cite{DaaGroVucMusKooSeiGieGor-JCP-20}
By linking $W_{c,\infty} $ and $W^{\rm DFT}_{\infty}$ and by knowing some exact features of $\beta[\rho^{\rm HF}]$,
Eq.~\eqref{eq:hfleq} will be used in this work for building approximations to $W_{c,\infty}$, exploiting the existing approximations for its DFT counterpart\cite{SeiPerKur-PRA-00,WagGor-PRA-14,BahZhoErn-JCP-16,VucGor-JPCL-17,GouVuc-JCP-2019}.

{\it Building approximations-} The AC framework has always played a crucial role in the construction of DFT approximations\cite{Bec-JCP-93a,Bec-JCP-93,PerErnBur-JCP-96,XYG3,ShaTouSav-JCP-11,LarGri-JCTC-10,SuXu-JCP-14,VucIroSavTeaGor-JCTC-16}, and, more recently, also in wave function theories, to approximate missing parts of the correlation energy (see, e.g., the work of Pernal and co-workers\cite{PasHapVeiPer-JPCL-2019,MarHapPerDeP-JCTC-2020}). In the present work, we build upon the interaction strength interpolation (ISI) idea of Seidl and co-workers,\cite{SeiPerLev-PRA-99,SeiPerKur-PRL-00} in which the DFT correlation energy is approximated by interpolating the AC integrand between its weak- and strong-coupling expansions. This construction enables one to include more pieces of information into the approximate correlation energy, avoiding a bias towards the weak correlation regime, present in most of the DFT approximations.\cite{SeiPerLev-PRA-99,SeiPerKur-PRL-00,VucIroWagTeaGor-PCCP-17,VucIroSavTeaGor-JCTC-16,VucGor-JPCL-17} The lack of size-consistency of the ISI approach had been considered its main drawback, but a recent remarkably simple size-consistency correction (SCC) fixes this problem in an exact way, at least for systems dissociating into fragments with a non-degenerate ground state.\cite{VucGorDelFab-JPCL-18,VucFabGorBur-JCTC-20} This SCC has been used for building  functionals\cite{VucGorDelFab-JPCL-18,VucFabGorBur-JCTC-20,Con-PRB-2019} in the DFT context, and has been already shown to be crucial for the accuracy of model MP AC curves (see Section S3 of Ref.~\citenum{VucFabGorBur-JCTC-20}) that can signal when an MP2 calculation is not reliable.\cite{VucFabGorBur-JCTC-20}

The idea of this work is to use this approach, originally designed for the DFT AC, in the MP AC context  to build accurate approximations to describe NCIs.
The mentioned SCC\cite{VucGorDelFab-JPCL-18} is also used throughout this work to restore size-consistency of our MP AC models.

\begin{figure}
\includegraphics[width=1\columnwidth]{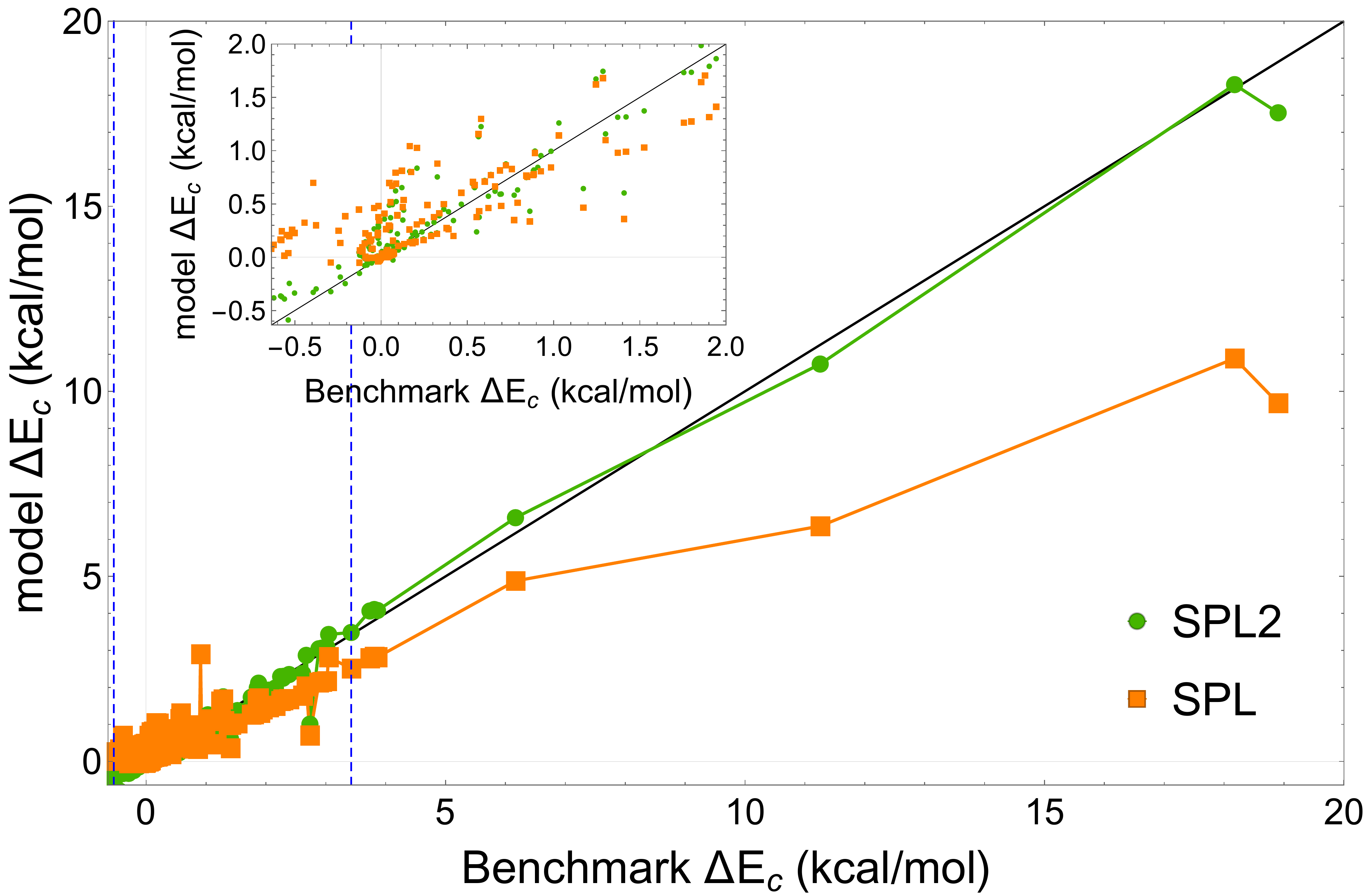}
\caption{Difference between benchmark CCSD(T) correlation energies and those of MP2 ($\Delta E_c = E_c - E_c^{\rm MP2}$) vs $\Delta E_c$ predicted by our models 
for a range of interaction energies of the NCI complexes [S22, S66, DI6, CT7, NGD8 and L7 datasets]. The blue vertical dashed lines denote the range in which $\Delta E_c$ values of complexes in S22, the dataset we used to train empirical parameters in the SPL2 form, lie.}
\label{fig:All_results}
\end{figure}

In a previous work on NCI's,\cite{VucGorDelFab-JPCL-18} the ISI idea has been applied within the DFT AC framework, by using MP2 as an approximation for the small-$\lambda$ expansion of the DFT AC. The interpolation function used was the one developed in the DFT AC framework by Seidl, Perdew and Levy (SPL),\cite{SeiPerLev-PRA-99}  
\beq
\label{eq:SPL}
 W_{c,\lambda}^{\rm SPL}=W_{c,\infty} 
\left (1 -  \frac{1}{\sqrt{1 + b \lambda} }\right) , 
\eeq
where 
$b=\left(4 E_{\rm c}^{\rm MP2}\right)/W_{c,\infty}$.
Such an attempt could be also viewed as an interpolation model for the MP AC in which the large-$\lambda$ limit was approximated by its DFT counterpart $W_{c,\infty}^{\rm DFT}$, which 
is known to be accurately described\cite{SeiGorSav-PRA-07,MirSeiGor-JCTC-12} by the point-charge plus continuum (PC) model\cite{SeiPerKur-PRA-00} 

\begin{align}
\label{eq:pc}
&W_{c,\infty} [\rho ^{\rm HF}]   \sim 
W_{c,\infty}^{\rm DFT} [\rho ^{\rm HF}]
\nonumber \\
&\approx \underbrace{ \int \left[A\rho^{\rm HF}(\mathbf{r})^{4/3} + B\frac{|\nabla \rho^{\rm HF}(\mathbf{r})|^2}{\rho^{\rm HF}(\mathbf{r})^{4/3}}\right]\mathrm{d}\mathbf{r} } _{W_{\infty}^{\rm PC}[\rho ^ {\rm HF}]} \nonumber \\
&-  E_{x}  [\{\phi_i^{\rm HF}\}],
\end{align}

where $A=-1.451$, $B=5.317\times10^{-3}$. Notice that this approximation is not in line with the exact relation of Eq.~\eqref{eq:hfleq}, which was not known at the time, but  it can still provide reasonable results. 
By performing a simple MP2 calculation and evaluating $W_{c,\infty}$ on the HF density, the needed quantities to be fed into Eq.~\eqref{eq:SPL} were easily obtained, yielding\cite{VucGorDelFab-JPCL-18,VucFabGorBur-JCTC-20} the SPL approximation to $E_{\rm c}$ (Eq.~\ref{eq:ac}). When it comes to 
NCIs, SPL was found\cite{VucGorDelFab-JPCL-18,VucFabGorBur-JCTC-20} to give a major improvement over MP2  (see Fig.~\ref{fig:All_results}). However, the deficiencies of SPL for NCIs are also already noticeable in Fig.~\ref{fig:All_results},
where we show results for various NCI's data sets. On the $x$-axis, we report the difference between the benchmark CCSD(T) and MP2 correlation energy  $\Delta E_c = E_c - E_c^{\rm MP2}$, and the difference between the benchmark and SPL correlation energies is reported on the $y$-axis. This way, if SPL had the same level of accuracy as the benchmark, all data points would lie on the $y=x$ line (shown in black). The plotted correlation energies pertain to the interaction energies, i.e. the differences between correlation energies of a complex and its fragments. Since MP2 (with large enough basis set) overbinds most of the complexes\cite{RezRilHob-JCTC-2011,VucFabGorBur-JCTC-20}, $\Delta E_c$ is positive for most of the datapoints. We can see from the same figure that 
SPL decently corrects MP2 for different ranges of $\Delta E_c$. As $\Delta E_c$ becomes large, SPL still substantially reduces the error of MP2. But, the performance of SPL is still not satisfactory as even the reduced errors can easily exceed 5 kcal/mol.  Moreover, for few systems where MP2 underbinds, SPL corrects it in the wrong direction (see the inset of Fig.~\ref{fig:All_results} that zooms in on the region around $\Delta E_c$=0). In these cases, a model MP AC integrand should be concave to correct the underbinding of MP2, and the SPL model is not  flexible enough to always capture this concavity\cite{VucFabGorBur-JCTC-20}. 

From these examples, it is clear that we need better interpolation forms than SPL to model $W_{c,\lambda}$, and a better description of its $\lambda\to\infty$ limit, which is known to be different\cite{SeiGiaVucFabGor-JCP-18,DaaGroVucMusKooSeiGieGor-JCP-20} than the DFT one used in this SPL construction. To make the interpolation form more flexible in capturing the  concavity/convexity correctly when MP2 underbinds/overbinds, respectively, we consider a form containing two SPL terms:
\beq \label {eq:spl2-M}
 W_{c,\lambda}^{\rm SPL2} = C_1 -  \frac{m_1}{\sqrt{1 + b_1 \lambda} } - \frac{m_2}{\sqrt{1 + b_2 \lambda} }.
\eeq
We call this form SPL2, with the $b_1$, $m_1$ and $C_1$ parameters fixed by the exact conditions: i) $W_{c,\lambda}^{\rm SPL2}$ vanishes at 0, ii) its initial derivative is equal to $2 E_c^{\rm MP2}$ (Eq.~\ref{eq:WHFMP}), and iii) it converges to  $W_{c,\infty}$ in the large $\lambda$ limit (Eq.~\ref{eq:large}),
\begin{eqnarray}
C_1 &= W^{\alpha\beta}_{c,\infty}; \nonumber \\
b_1 &= \frac{b_2~m_2- 4 E_c^{\rm MP2}}{m_2 - W^{\alpha\beta}_{c,\infty}}; \nonumber \\
m_1 &= W^{\alpha\beta}_{c,\infty} - m_2,
\end{eqnarray}
where $W^{\alpha\beta}_{c,\infty}  [\rho^{\rm HF}]$ is an approximation to $W_{c,\infty}  [\rho^{\rm HF}]$ inspired by Eq.~\eqref{eq:hfleq}:
\begin{align}\label{eq:ab}
 W^{\alpha\beta}_{c,\infty}  [\rho^{\rm HF}]=  \alpha W^{\rm PC}_{\infty} [\rho^{\rm HF}] + \beta E_{x}  [\{\phi_i^{\rm HF}\}]. 
\end{align} 

In fact, as mentioned, the exact form of $W_{c,\infty}$ for the MP AC has been recently revealed,\cite{SeiGiaVucFabGor-JCP-18,DaaGroVucMusKooSeiGieGor-JCP-20} but it is quite involved, with targeted semilocal approximations still under construction; the use of Eq.~\eqref{eq:hfleq} to improve the large-$\lambda$ description of the MP AC seems a rather effective first step.
There are now 4 parameters left in the SPL2 model for MP AC ($b_2$, $m_2$, $\alpha$, and $\beta$ that we simplify to be system independent),
which we fit in this work to the S22 dataset\cite{S22,S22ref} by minimizing its mean absolute error (MAE). From Fig.~\ref{fig:All_results}, we can see that SPL2 fixes the key deficiencies of SPL: it corrects MP2 in the right direction when the latter underbinds, and has a better corrective trend than SPL as MP2 errors become large. 
\par
In addition to the SPL2 model, we 
also develop a model for $E_{c}$ directly. We first generalize $E_{\rm c}$ as:
$E_{\rm c,\lambda}= \int_0^\lambda W_{\rm c,\lambda'} \dd \lambda'$, such that
$E_{\rm c}=E_{\rm c,\lambda = 1}$. Then we build the following model for  $E_{\rm c,\lambda}$: 
\begin{equation}
E_{\rm c,\lambda}^{\rm MPACF-1}=-g \lambda +\frac{g (h+1) \lambda}{\sqrt{d_1^2 \lambda +1}+h \sqrt[4]{d_2^4 \lambda+1}},
\end{equation}
with
\begin{eqnarray}
    g&= -W^{1, 1}_{c,\infty}\nonumber \\
    h&= \frac{\displaystyle 4E_{c}^{\rm MP2} - 2 d_1^2 W^{1, 1}_{c,\infty}}{\displaystyle -4 E_{c}^{\rm MP2} + d_2^4 W^{1,1}_{c,\infty}},
\end{eqnarray}
where $W^{1, 1}_{c,\infty}$ is  $W^{\alpha\beta}_{c,\infty}$ in which $\alpha$ and $\beta$ are set to $1$ by using the UEG argument (see above). This new model is called Møller-Plesset Adiabatic Connection Functional-1 (MPACF-1) and will be the starting point for a new class of functionals that approximate the MPAC.
The underlying MP AC model, $W_{\rm c,\lambda}^{\rm MPACF-1}$, is simply obtained by taking a derivative of $E_{\rm c,\lambda}^{\rm MPACF-1}$ w.r.t. $\lambda$.
In contrast to  $W_{\rm c,\lambda}^{\rm SPL}$ and $W_{\rm c,\lambda}^{\rm SPL2}$, $W_{\rm c,\lambda}^{\rm MPACF-1}$ contains the $W_{\frac{3}{4}}$ term appearing in the large $\lambda$ limit (Eq.~\ref{eq:large}), making this model having a better asymptotic behavior than SPL and SPL2. 
MPACF-1 is also less empirical than SPL2 since it contains only two parameters ($d_1$ and $d_2$),  which we again fit to the S22 datasets and report their optimal values in the Computational Details below.

Without the SCC, SPL2 and MPACF-1 have a different size-extensivity behaviour. Nevertheless, in this work we always use the SCC ensuring that the interaction energies are correctly computed and vanish in the dissociation limit (see Fig. S1 from the Supplementary Information\cite{sup} for the Kr$_2$ example and the caption of this figure for a further discussion). From Fig. S1, one can also see that the SCC does not affect the shape of the potential energy surfaces (PES), but only shifts a PES by a constant ensuring that binding energies vanish when the fragments are infinitely away from one another. Thus, the SCC would not be required and has no effect for calculating differences in energies at different stationary points of a PES (e.g. reaction energies, barrier heights, isomerisation energies, etc).

In what follows, we compare the performance of SPL2 and MPACF-1 with that of earlier SPL, MP2 and other approximations for NCIs. While our models share some similarities with DHs, there are two key differences. First, our models are based on the full amounts of the exact exchange and MP2 correlation and 
thus they do not benefit from error cancellations between these quantities and their semilocal counterparts. Furthermore, our models do not require dispersion corrections to be accurate for NCIs. 

 \begin{figure}
\includegraphics[width=1\columnwidth]{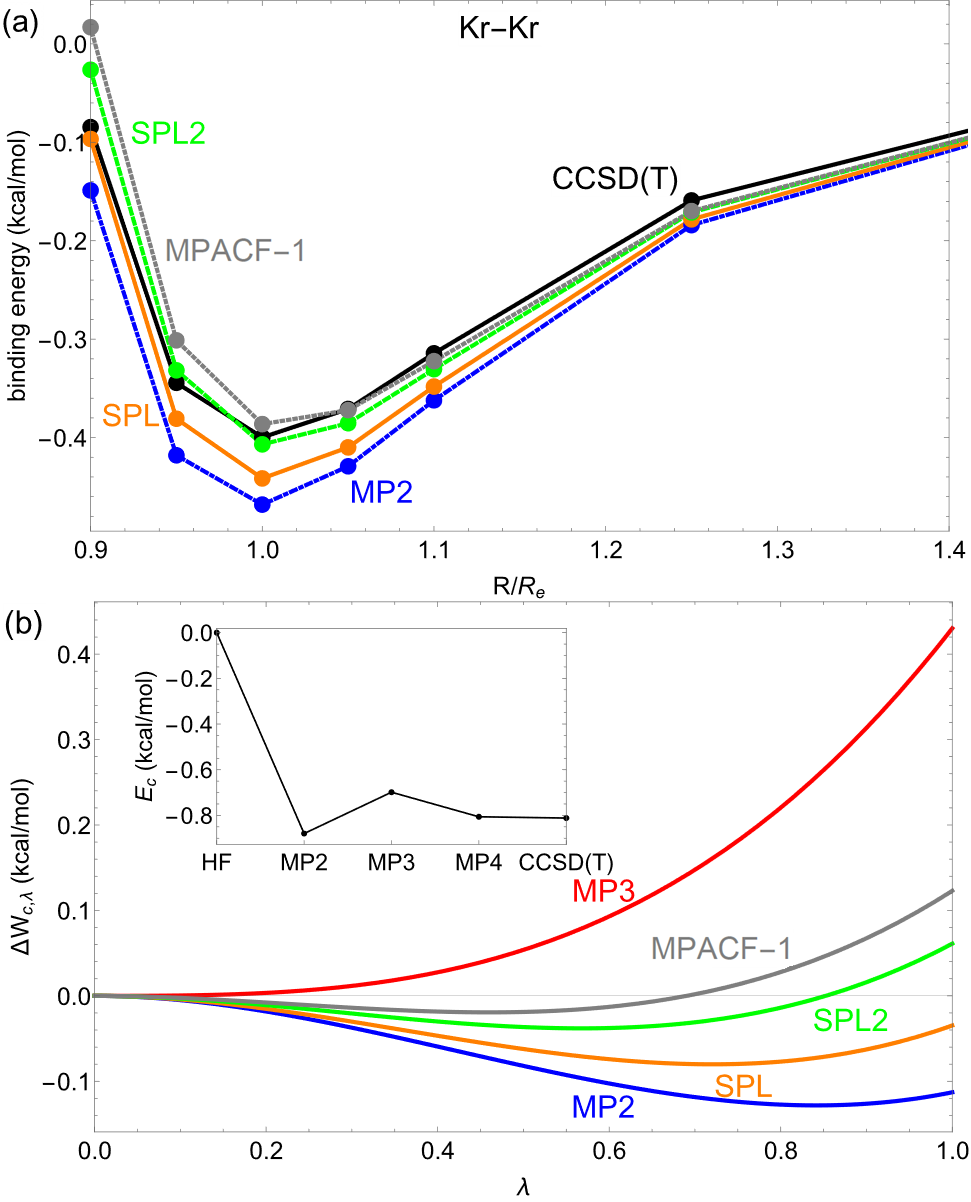}
\caption{Panel (a): The interaction energies of MP2, SPL, SPL2 and MPACF-1 as well as reference CCSD(T) curves for Kr$_2$. 
Panel (b): The errors of different MP AC models for Kr$_2$ at equilibrium, $\Delta W_{c,\lambda} = W_{c,\lambda}^{\rm method} -W_{c,\lambda} ^{\rm ref}$, where the
 r.h.s. of Eq~\ref{eq:WHFMP} truncated to fourth order, $W_{c,\lambda} ^{\rm MP4}$, is taken as a reference (the inset justifies this choice given the (fast) convergence of MP$n$ series).}
\label{fig:kr2}
\end{figure}

{\it Results-} We start with a light example, showing the Kr$_2$ binding curve  in Fig.~\ref{fig:kr2}. 
We can see from the top panel of Fig.~\ref{fig:kr2}
that SPL significantly improves MP2.
At the same time,  SPL is significantly improved by MPACF-1 and SPL2, with the latter being slightly more accurate than the former. 
This happens even though 
noble gas dimers are beyond our training set (S22).
In contrast, 
the D3 empirical correction can even worsen binding curves of noble gas dimers,
\cite{VucBur-JPCL-20} even though their binding energies have been used in the training of the original D3 parameters\cite{GAEK10}. 
Now we move to the bottom panel of Fig.~\ref{fig:kr2}, where we look at the accuracy of different AC models for the interaction energies for Kr$_2$ at equilibrium. To test the accuracy of our AC models, we need a reference $W_{c,\lambda}$ for the interaction energies. 
Ideally, this quantity would be obtained by full-CI or CCSD(T), but we note that the convergence of MP$n$ series for the interaction energy of Kr$_2$ at equilibrium
is fast (see the inset of the lower panel of Fig.~\ref{fig:kr2} showing that MP4 gives nearly the same results as CCSD(T)). 
For this reason, we can safely assume that 
the r.h.s. of Eq.~\eqref{eq:WHFMP} truncated to fourth order gives us a reliable MP AC reference for the interaction energies of Kr$_2$. After establishing $W_{c,\lambda} ^ {\rm MP4}$ as a reference, we compare the performance of MP3, MP2, SPL, SPL2 and MPACF-1 curves in the lower panel of Fig.~\ref{fig:kr2}). 
The error of all MP AC models slowly increases as we move away from $\lambda=0$ since all the curves have the correct initial slope given by $2E_{\rm c}^{\rm MP2}$. 
On average, SPL2 is the most accurate. MPACF-1 is the most accurate up to $\lambda \sim 0.7$, then its accuracy deteriorates and at about $\lambda \sim 0.9$ where it is less accurate than even SPL.  Overall, all three AC models give significant improvements over MP2 and MP3.

\begin{table}[!htb]
\caption{The MAE  in kcal/mol of different methods for the S22, CT7, DI6, S66 and L7 datasets from the existing literature. Best results are highlighted in bold. NGD8 is a set of 8 noble gas dimers (\ce{Ar2}, \ce{He2}, \ce{Kr2}, \ce{Ne2}, \ce{ArKr}, \ce{C6H6-Ne}, \ce{CH4Ne} and \ce{HeAr}) that we construct here.}
\begin{tabular}{llllll}
 \multicolumn{1}{l|}{set} & \text{MP2} & \text{SPL} & \text{SPL2} & \text{MPACF-1} & \text{B3LYP-D3}\\ \cline{1-6}
 \multicolumn{1}{l|}{NGD8} & 0.04 & 0.05 & {\bf 0.03} & {\bf 0.03} & 0.08 \\
 \multicolumn{1}{l|}{CT7} & 0.92 & 0.57 & {\bf 0.45} & 0.60 & 1.48\\
 \multicolumn{1}{l|}{DI6} & 0.48 & 0.27 & {\bf 0.18} & 0.20 & 0.46\\
 \multicolumn{1}{l|}{S22} & 0.88 & 0.38 & {\bf 0.15} & 0.19 & {\bf 0.15}\\
 \multicolumn{1}{l|}{S66} & 0.47 & 0.35 & 0.21 & 0.26 & {\bf 0.18}\\
 \multicolumn{1}{l|}{L7} & 8.74 & 3.83 & {\bf 0.89}  & 2.32 & 1.78
\label{tab_results}
\end{tabular}
\end{table}

\begin{figure*}
\includegraphics[width=0.95\textwidth]{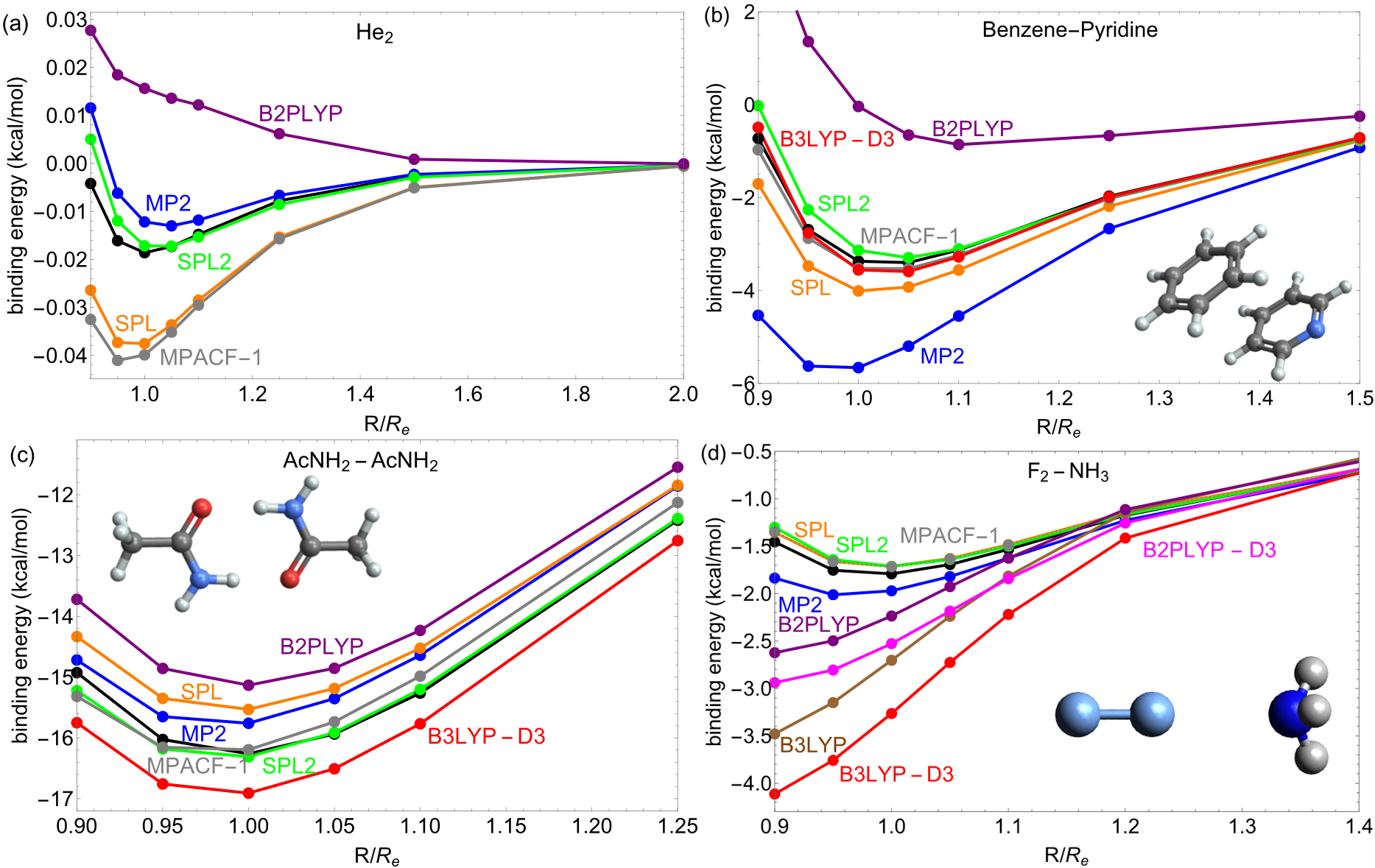}
\caption{Dissociation curves of  He$_2$ (a), benzene-pyridine (b), acetamide (c), and \ce{NH3}-\ce{F2} dimers (d), obtained from various methods. CCSD(T) (black line) has been used as a reference for all complexes. For other binding curves, see Figs. S3-S8 from the Supplementary Information\cite{sup}.
}
\label{fig:cur}
\end{figure*}

Now we move to Table~\ref{tab_results}, where we show the results for several datasets \cite{S22,RezRilHob-JCTC-2011,ZhaTru-JCTC-2005,ZhaSchTru-JCP-2005,ZhaSchTru-JCTC-06,SedJanPitRezPulHob-ACS-2013} in comparison with the B3LYP hybrid enhanced by D3. Except for noble gas dimers where the differences in MAEs are marginal, SPL improves the performance of MP2 by a factor of $2$ on average. SPL2 greatly improves SPL by reducing its errors by the factors ranging from $1.3$ (CT7) to more than $4$ (L7). 
MPACF-1 provides still a substantial improvement over SPL, but on average performs worse than SPL2. 
SPL2 also beats B3LYP-D3 for NGD8, CT7, DI6 and L7, whereas the two approaches display a similar performance for S22 and S66. 
By looking at MAEs for individual subsets of the S66 dataset (see Table S1 from the Supplementary Information\cite{sup}),
we can  see that the accuracy of MP2 is high for hydrogen bonded complexes, but it is well-known that it  deteriorates for dispersion-bonded and mixed complexes\cite{RezRilHob-JCTC-2011,VucFabGorBur-JCTC-20}. SPL greatly reduces the errors of MP2 for dispersion-bonded and mixed complexes, but becomes worse than MP2 for hydrogen bonds. On the other hand, SPL2 greatly improves MP2 for dispersion-bonded and mixed complexes, without deteriorating its accuracy for hydrogen bonds.

In Fig.~\ref{fig:cur}, we show several binding curves representing NCIs of different nature. 
This includes weakly-bonded He$_2$ [panel (a)], stacked Benzene-Pyridine complex [panel (b)], hydrogen-bonded acetamide dimer [panel (c)], and the charged-transfer (CT) fluorine-ammonia complex [panel (d)].
Overall,  SPL2 is in the closest agreement with the reference [CCSD(T)] and it corrects MP2 in the right direction in cases when MP2 underbinds (He$_2$, the acetamide dimer), when it overbinds slightly (the fluorine-ammonia complex), and 
severely (the benzene-pyridine complex).
On the other hand, SPL corrects MP2 in the right direction
only in cases when the latter overbinds. MPACF-1 is off for He$_2$, but in other cases it is on par with SPL2 and  even more accurate than SPL2 in the case of the benzene-pyridine dimer. For He$_2$, SPL is as bad as MPACF-1, whereas empirical SPL2 is very accurate (even though noble gases have not been used in the training of SPL2). Thus, it seems indeed challenging to build non-empirical MP AC model that will give improvements over MP2 for He$_2$, but may be achieved in the future by considering approximations to  higher-order terms from the large $\lambda$ limit of the MP AC.\cite{DaaGroVucMusKooSeiGieGor-JCP-20}

Our models are accurate for NCIs without requiring dispersion corrections (in contrast to e.g., D3-uncorrected B2PLYP, which is off for all four cases).
B3LYP corrected by D3 is completely off for He$_2$ (see Fig. S3 from the Supplementary Information\cite{sup}).
The behavior of B3LYP and B2PLYP is even more interesting in the case of the CT fluorine-ammonia complex where the D3 correction even worsens the original results\cite{ZhaXuJunGod-PNAS-2011}. This is not due to the D3 correction itself, but due to the density-driven errors that typically bedevil semilocal DFT calculations of halogen-based CT complexes \cite{KSSB18,SonVucSimBur-JPCL-21,lars2021} (Note also that is not uncommon that the D3 correction worsens the results from semilocal DFT calculations suffering from density-driven errors\cite{KSSB18,SonVucSimBur-JPCL-21}). On the contrary, our MP AC models are built only for correlation, and thus the full amount of the exact exchange is used without being mixed with a fraction of its semilocal counterpart. This is probably the reason why all of our MP AC models are very accurate for the studied CT complex (see also Fig. S7 from the Supplementary Information\cite{sup} for another example).

\begin{figure}
\includegraphics[width=1\columnwidth]{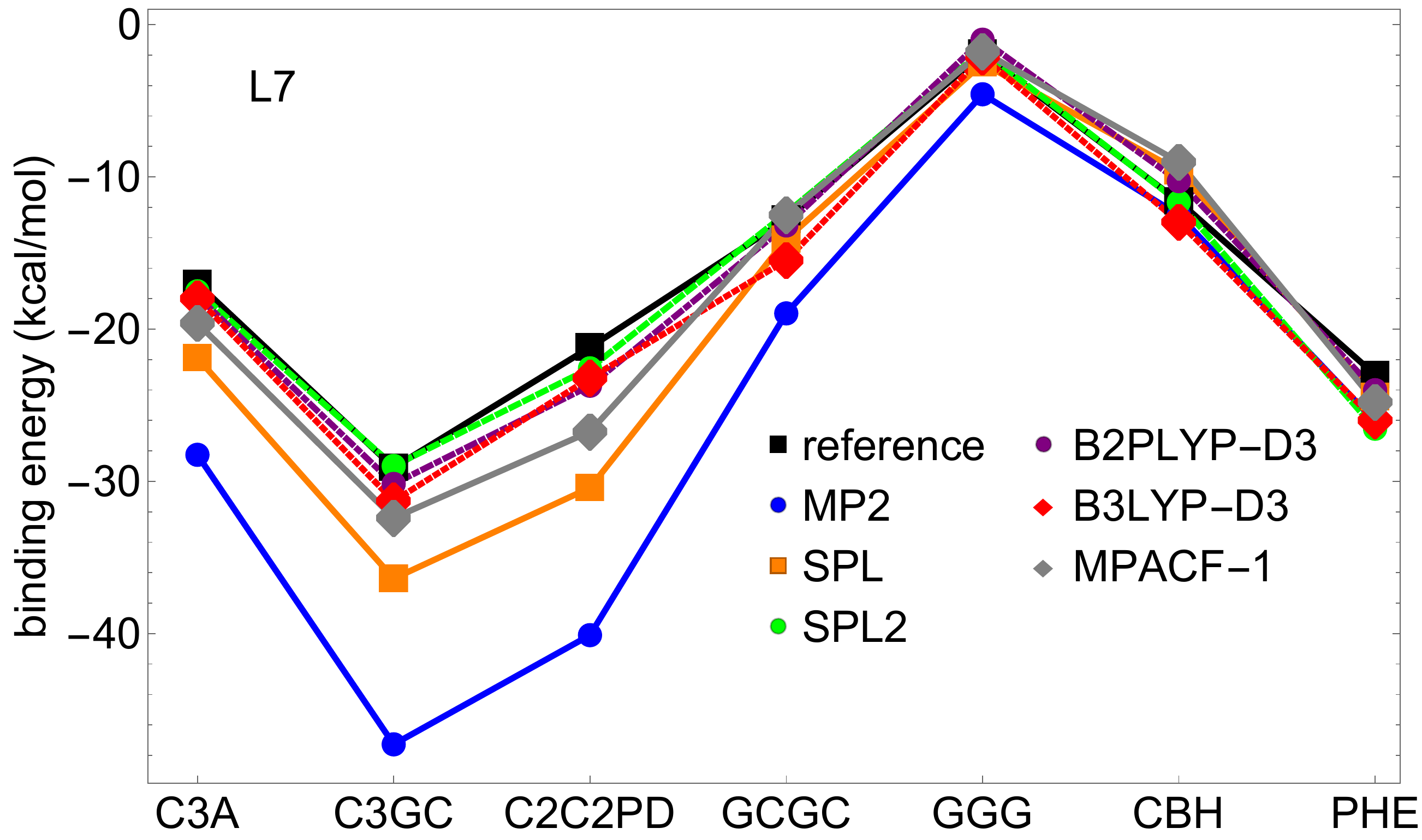}
\caption{Interaction energies of MP2, SPL, SPL2, B3LYP-D3 and B2PLYP as well as reference data of Grimme and co-workers\cite{GriBraBanHan-JCP-2015}, plotted for individual complexes of the L7 dataset. For comparison of the approximations against other reference data, see Fig.~S2 and Table~S2 from the Supplementary Information\cite{sup}.}
\label{fig:L7}
\end{figure}

Now we go back to the L7 dataset composed by larger complexes for which MP2 displays very large errors\cite{Ngu-Fur-JCTC-2020}. Interaction energies for individual L7 complexes are shown in Fig.~\ref{fig:L7}, where the reference used is obtained from Grimme et al.\cite{GriBraBanHan-JCP-2015}.
From Fig.~\ref{fig:L7}, we can see that MP2 strongly overbinds most the L7 complexes. SPL corrects it, but not sufficiently, as it is still much less accurate than B3LYP-D3 and B2PLYP-D3. MPACF-1 improves SPL, but a better performance is obtained from our SPL2 model, which very accurately reproduces the reference values. In the SI (Table S2 and Fig. S2 from the Supplementary Information\cite{sup}), we compare the performance of our models against other L7 references from the literature.\cite{SedJanPitRezPulHob-ACS-2013,HamNagBarKalBraTka-unknown-2020} From these results, it can be seen that regardless of what reference is used, SPL greatly reduces the MP2 errors, whereas our new models (SPL2 and MPACF-1) greatly reduce the errors of SPL.

{\it Conclusions and Perspectives --} In summary, we have introduced a new scheme for the construction of MP AC models providing accurate description of NCIs. Two specific interpolation models,  SPL2 and MPACF-1 greatly reduce errors of MP2 and earlier proposed SPL for a variety of NCIs despite coming at a negligible computational cost beyond that of MP2. 
In comparison with e.g. modern (double) hybrids\cite{MH16,MH17,MHG18}, empirical parameters in our models have been primitively optimized and despite that offer highly competitive accuracy for the description of NCIs. Further improvements can be obtained by better optimization strategies, but also by reducing empiricism using the additional recently revealed exact form of  the large $\lambda$ limit MP AC functionals. \cite{DaaGroVucMusKooSeiGieGor-JCP-20} 
Furthermore, one can use machine learning to find improved ways for interpolating MP AC (see, e.g. Ref~\citenum{SNSMP2} for a related work). 
To lower the cost, in future work we will re-design our models by replacing the exact $E_{\rm c}^{\rm MP2}$ with some of its approximations. \cite{SHW99,LMH00,WilWilMan-JCTC-2020}
The zero-cost size-consistency correction of Ref.~\onlinecite{VucGorDelFab-JPCL-18} is an important part of our scheme and for now it can be applied only to systems dissociating into non-degenerate ground states. Generalization of this correction for systems that dissociate into degenerate ground state fragments  will be another objective for future works. This and the investigation of the large-coupling limit of the MP AC for open-shell systems should pave the way for the broader applicability of our scheme (i.e. beyond NCIs). The development of analytical gradients enabling further applicability of our method will be then considered following the similar
implementation of gradients for standard double hybrids.\cite{grime}. Our scheme can also be used to give a formal justification and to  improve recently introduced double-hybrid functionals that are applied to Hartree-Fock densities\cite{SonVucSimBur-JPCL-21}. 

\section{Computational details}

All calculations have been performed using a modified version of the TURBOMOLE~7.1 package \cite{Furche2014,TURBOMOLE}. Computational details are the same as in Refs. \citenum{FabGorSeiDel-JCTC-16,GiaGorDelFab-JCP-18}. For all MP2 calculations we have employed aug-cc-pVQZ enhanced with additional basis functions detailed in Ref.~\citenum{VucGorDelFab-JPCL-18}. MP2 interaction energies for NCIs within this basis set are close to the MP2/CBS results\cite{VucFabGorBur-JCTC-20}. From these MP2 calculations we extract all the quantities needed to construct our MP AC interpolation models (HF densities, $E_{x}  [\{\phi_i^{\rm HF}\}]$, $E_{\rm c}^{\rm MP2}$). The B3LYP-D3 results shown in Table~\ref{tab_results} for the S22, CT7 DI6, S66, and L7 datasets have been taken from Refs.~\citenum{GriBraBanHan-JCP-2015,ZhaXuJunGod-PNAS-2011,ZhaXuJunGod-PNAS-2011,GoeHanBauEhrNajGri-RSC-2017,SedJanPitRezPulHob-ACS-2013}, respectively, whereas for NGD8 they were calculated using an aug-cc-pVQZ basis set. The same basis set has been employed to calculate MP3 and MP4 energies for Kr$_2$.
For B2PLYP-D3 the data for L7 was obtained from Ref\citenum{CalOrtSanAra-JCTC-2015}. The B2PLYP, B3LYP-D3 and CCSD(T)/CBS data for the S66 dissociation curves were obtained from~\citenum{RezRilHob-JCTC-2011}. The B2PLYP(-D3), B3LYP(-D3) and CCSD(T)/CBS data for the CT7 dissociation curves were calculated using an aug-cc-pVQZ basisset. The B3LYP-D3 for the dissociation curves of \ce{Ne2}, \ce{Kr2} and \ce{Ar2} were obtained form \citenum{KovDobOstRod-2017-IJQC-2017}, whereas the rest of the B3LYP-D3, B2PLYP and CCSD(T) data were again calculated with the basisset aug-cc-pVQZ.

For SPL2 the optimal parameters we find are ($b_2=0.117$, $m_2=10.68$, $\alpha=1.1472$, $\beta=-0.7397$), whereas for MPACF-1 we use the following set of parameters, ($d_1=0.294$, $d_2=0.934$). A few remarks on these are parameters are needed. As shown in Ref.~\citenum{DaaGroVucMusKooSeiGieGor-JCP-20}, the large-$\lambda$ limit of the MP AC integrand $W_{\rm c,\lambda}$ has a leading term $W_{\rm c,\infty}$ which is much lower than its DFT counterpart. At the next order, the $\lambda^{-1/2}$ term is, instead, positive and much larger than its DFT counterpart, because the HF exchange operator enhances the zero-point energy, by introducing excited states of the normal modes. \cite{DaaGroVucMusKooSeiGieGor-JCP-20} Finally, the $\lambda^{-3/4}$ term is not present in the DFT AC, and it is a peculiar feature of the MP AC.\cite{DaaGroVucMusKooSeiGieGor-JCP-20} This term is again negative. Overall, these three terms together are needed for an accurate description of $W_{\rm c,\lambda}$, as they balance each other in a delicate way (see Fig.~9 of Ref.~\citenum{DaaGroVucMusKooSeiGieGor-JCP-20}). The SPL2 form does not have the $\lambda^{-3/4}$ term. For this reason, its large-$\lambda$ limit $W_{\rm c,\infty}$ is an effective description of the three leading terms; this is why in this case $\beta$ turns out to be negative. In addition, we need to point out that the fitting procedure do not consider the total energies but takes into account only the interaction energies and should {\bf only} be used
within the SCC approach. The new MPACF-1 form, instead, has built in the correct large-$\lambda$ behavior, including the   $\lambda^{-3/4}$ term. This is why in this case the parameters $\alpha$ and $\beta$ can be set equal to 1. In future work we will build accurate GGA functionals for the first two leading terms of the large-$\lambda$ MP AC, which are expected to improve our models, when combined with an approximation for the $\lambda^{-3/4}$ term containing the HF density at the nuclei. 
In Table S3 from the Supplementary Information\cite{sup}, we  give a summary of the three MP AC forms, their parameters, etc., and highlight the differences in models for $W_{c,\infty}$ that our three forms use.

\section{Supporting Information} 
The Supporting Information is available free of charge on the ACS Publications website ... at DOI: ... 
\begin{itemize}
    \item {MAEs for S66 subsets, additional dissociation curves of noncovalent complexes, plots comparing errors for the L7 dataset, summary of MP AC forms developed in this work. }
\end{itemize}

\section{Acknowledgments} 
Financial support by the Netherlands Organisation for Scientific Research under Vici grant 724.017.001 is acknowledged.
This work was also supported the European Research Council under H2020/ERC Consolidator Grant corr-DFT (Grant No. 648932).
SV acknowledges financial support from the Alexander von Humboldt Foundation. 

\bibliography{bib_clean,sh8}

\providecommand{\latin}[1]{#1}
\makeatletter
\providecommand{\doi}
  {\begingroup\let\do\@makeother\dospecials
  \catcode`\{=1 \catcode`\}=2 \doi@aux}
\providecommand{\doi@aux}[1]{\endgroup\texttt{#1}}
\makeatother
\providecommand*\mcitethebibliography{\thebibliography}
\csname @ifundefined\endcsname{endmcitethebibliography}
  {\let\endmcitethebibliography\endthebibliography}{}
\begin{mcitethebibliography}{95}
\providecommand*\natexlab[1]{#1}
\providecommand*\mciteSetBstSublistMode[1]{}
\providecommand*\mciteSetBstMaxWidthForm[2]{}
\providecommand*\mciteBstWouldAddEndPuncttrue
  {\def\EndOfBibitem{\unskip.}}
\providecommand*\mciteBstWouldAddEndPunctfalse
  {\let\EndOfBibitem\relax}
\providecommand*\mciteSetBstMidEndSepPunct[3]{}
\providecommand*\mciteSetBstSublistLabelBeginEnd[3]{}
\providecommand*\EndOfBibitem{}
\mciteSetBstSublistMode{f}
\mciteSetBstMaxWidthForm{subitem}{(\alph{mcitesubitemcount})}
\mciteSetBstSublistLabelBeginEnd
  {\mcitemaxwidthsubitemform\space}
  {\relax}
  {\relax}

\bibitem[Hohenstein and Sherrill(2012)Hohenstein, and Sherrill]{hohenstein12}
Hohenstein,~E.~G.; Sherrill,~C.~D. Wavefunction methods for noncovalent
  interactions. \emph{WIREs Computational Molecular Science} \textbf{2012},
  \emph{2}, 304--326\relax
\mciteBstWouldAddEndPuncttrue
\mciteSetBstMidEndSepPunct{\mcitedefaultmidpunct}
{\mcitedefaultendpunct}{\mcitedefaultseppunct}\relax
\EndOfBibitem
\bibitem[Riley and Hobza(2011)Riley, and Hobza]{riley11}
Riley,~K.~E.; Hobza,~P. Noncovalent interactions in biochemistry. \emph{WIREs
  Computational Molecular Science} \textbf{2011}, \emph{1}, 3--17\relax
\mciteBstWouldAddEndPuncttrue
\mciteSetBstMidEndSepPunct{\mcitedefaultmidpunct}
{\mcitedefaultendpunct}{\mcitedefaultseppunct}\relax
\EndOfBibitem
\bibitem[Lao and Herbert(2015)Lao, and Herbert]{lao15}
Lao,~K.~U.; Herbert,~J.~M. Accurate and Efficient Quantum Chemistry
  Calculations for Noncovalent Interactions in Many-Body Systems: The XSAPT
  Family of Methods. \emph{The Journal of Physical Chemistry A} \textbf{2015},
  \emph{119}, 235--252\relax
\mciteBstWouldAddEndPuncttrue
\mciteSetBstMidEndSepPunct{\mcitedefaultmidpunct}
{\mcitedefaultendpunct}{\mcitedefaultseppunct}\relax
\EndOfBibitem
\bibitem[Sedlak \latin{et~al.}(2013)Sedlak, Janowski, Pitoňák, Řezáč,
  Pulay, and Hobza]{sedlak13}
Sedlak,~R.; Janowski,~T.; Pitoňák,~M.; Řezáč,~J.; Pulay,~P.; Hobza,~P.
  Accuracy of Quantum Chemical Methods for Large Noncovalent Complexes.
  \emph{Journal of Chemical Theory and Computation} \textbf{2013}, \emph{9},
  3364--3374\relax
\mciteBstWouldAddEndPuncttrue
\mciteSetBstMidEndSepPunct{\mcitedefaultmidpunct}
{\mcitedefaultendpunct}{\mcitedefaultseppunct}\relax
\EndOfBibitem
\bibitem[Gráfová \latin{et~al.}(2010)Gráfová, Pitoňák, Řezáč, and
  Hobza]{grafova10}
Gráfová,~L.; Pitoňák,~M.; Řezáč,~J.; Hobza,~P. Comparative Study of
  Selected Wave Function and Density Functional Methods for Noncovalent
  Interaction Energy Calculations Using the Extended S22 Data Set.
  \emph{Journal of Chemical Theory and Computation} \textbf{2010}, \emph{6},
  2365--2376\relax
\mciteBstWouldAddEndPuncttrue
\mciteSetBstMidEndSepPunct{\mcitedefaultmidpunct}
{\mcitedefaultendpunct}{\mcitedefaultseppunct}\relax
\EndOfBibitem
\bibitem[Al-Hamdani and Tkatchenko(2019)Al-Hamdani, and
  Tkatchenko]{alhamdani19}
Al-Hamdani,~Y.~S.; Tkatchenko,~A. Understanding non-covalent interactions in
  larger molecular complexes from first principles. \emph{The Journal of
  Chemical Physics} \textbf{2019}, \emph{150}, 010901\relax
\mciteBstWouldAddEndPuncttrue
\mciteSetBstMidEndSepPunct{\mcitedefaultmidpunct}
{\mcitedefaultendpunct}{\mcitedefaultseppunct}\relax
\EndOfBibitem
\bibitem[Grimme \latin{et~al.}(2016)Grimme, Hansen, Brandenburg, and
  Bannwarth]{grimme16}
Grimme,~S.; Hansen,~A.; Brandenburg,~J.~G.; Bannwarth,~C. Dispersion-Corrected
  Mean-Field Electronic Structure Methods. \emph{Chemical Reviews}
  \textbf{2016}, \emph{116}, 5105--5154\relax
\mciteBstWouldAddEndPuncttrue
\mciteSetBstMidEndSepPunct{\mcitedefaultmidpunct}
{\mcitedefaultendpunct}{\mcitedefaultseppunct}\relax
\EndOfBibitem
\bibitem[Dubecký \latin{et~al.}(2016)Dubecký, Mitas, and Jurečka]{dubecky16}
Dubecký,~M.; Mitas,~L.; Jurečka,~P. Noncovalent Interactions by Quantum Monte
  Carlo. \emph{Chemical Reviews} \textbf{2016}, \emph{116}, 5188--5215\relax
\mciteBstWouldAddEndPuncttrue
\mciteSetBstMidEndSepPunct{\mcitedefaultmidpunct}
{\mcitedefaultendpunct}{\mcitedefaultseppunct}\relax
\EndOfBibitem
\bibitem[Christensen \latin{et~al.}(2016)Christensen, Kubař, Cui, and
  Elstner]{christensen16}
Christensen,~A.~S.; Kubař,~T.; Cui,~Q.; Elstner,~M. Semiempirical Quantum
  Mechanical Methods for Noncovalent Interactions for Chemical and Biochemical
  Applications. \emph{Chemical Reviews} \textbf{2016}, \emph{116},
  5301--5337\relax
\mciteBstWouldAddEndPuncttrue
\mciteSetBstMidEndSepPunct{\mcitedefaultmidpunct}
{\mcitedefaultendpunct}{\mcitedefaultseppunct}\relax
\EndOfBibitem
\bibitem[Burns \latin{et~al.}(2011)Burns, Mayagoitia, Sumpter, and
  Sherrill]{burns11}
Burns,~L.~A.; Mayagoitia,~A.~V.; Sumpter,~B.~G.; Sherrill,~C.~D.
  Density-functional approaches to noncovalent interactions: A comparison of
  dispersion corrections (DFT-D), exchange-hole dipole moment (XDM) theory, and
  specialized functionals. \emph{The Journal of Chemical Physics}
  \textbf{2011}, \emph{134}, 084107\relax
\mciteBstWouldAddEndPuncttrue
\mciteSetBstMidEndSepPunct{\mcitedefaultmidpunct}
{\mcitedefaultendpunct}{\mcitedefaultseppunct}\relax
\EndOfBibitem
\bibitem[DiLabio \latin{et~al.}(2013)DiLabio, Johnson, and Otero-de-la
  Roza]{dilabio13}
DiLabio,~G.~A.; Johnson,~E.~R.; Otero-de-la Roza,~A. Performance of
  conventional and dispersion-corrected density-functional theory methods for
  hydrogen bonding interaction energies. \emph{Phys. Chem. Chem. Phys.}
  \textbf{2013}, \emph{15}, 12821--12828\relax
\mciteBstWouldAddEndPuncttrue
\mciteSetBstMidEndSepPunct{\mcitedefaultmidpunct}
{\mcitedefaultendpunct}{\mcitedefaultseppunct}\relax
\EndOfBibitem
\bibitem[Grimme \latin{et~al.}(2010)Grimme, Antony, Ehrlich, and Krieg]{GAEK10}
Grimme,~S.; Antony,~J.; Ehrlich,~S.; Krieg,~H. A consistent and accurate ab
  initio parametrization of density functional dispersion correction (DFT-D)
  for the 94 elements H-Pu. \emph{The Journal of Chemical Physics}
  \textbf{2010}, \emph{132}, 154104\relax
\mciteBstWouldAddEndPuncttrue
\mciteSetBstMidEndSepPunct{\mcitedefaultmidpunct}
{\mcitedefaultendpunct}{\mcitedefaultseppunct}\relax
\EndOfBibitem
\bibitem[Hobza(2011)]{hobza11}
Hobza,~P. The calculation of intermolecular interaction energies. \emph{Annu.
  Rep. Prog. Chem.{,} Sect. C: Phys. Chem.} \textbf{2011}, \emph{107},
  148--168\relax
\mciteBstWouldAddEndPuncttrue
\mciteSetBstMidEndSepPunct{\mcitedefaultmidpunct}
{\mcitedefaultendpunct}{\mcitedefaultseppunct}\relax
\EndOfBibitem
\bibitem[Laricchia \latin{et~al.}(2012)Laricchia, Fabiano, and {Della
  Sala}]{laricchia12}
Laricchia,~S.; Fabiano,~E.; {Della Sala},~F. On the accuracy of frozen density
  embedding calculations with hybrid and orbital-dependent functionals for
  non-bonded interaction energies. \emph{The Journal of Chemical Physics}
  \textbf{2012}, \emph{137}, 014102\relax
\mciteBstWouldAddEndPuncttrue
\mciteSetBstMidEndSepPunct{\mcitedefaultmidpunct}
{\mcitedefaultendpunct}{\mcitedefaultseppunct}\relax
\EndOfBibitem
\bibitem[Grabowski \latin{et~al.}(2013)Grabowski, Fabiano, and {Della
  Sala}]{grabowski13}
Grabowski,~I.; Fabiano,~E.; {Della Sala},~F. A simple non-empirical procedure
  for spin-component-scaled MP2 methods applied to the calculation of the
  dissociation energy curve of noncovalently-interacting systems. \emph{Phys.
  Chem. Chem. Phys.} \textbf{2013}, \emph{15}, 15485--15493\relax
\mciteBstWouldAddEndPuncttrue
\mciteSetBstMidEndSepPunct{\mcitedefaultmidpunct}
{\mcitedefaultendpunct}{\mcitedefaultseppunct}\relax
\EndOfBibitem
\bibitem[Fabiano \latin{et~al.}(2014)Fabiano, Constantin, and {Della
  Sala}]{fabiano14}
Fabiano,~E.; Constantin,~L.~A.; {Della Sala},~F. Wave Function and Density
  Functional Theory Studies of Dihydrogen Complexes. \emph{Journal of Chemical
  Theory and Computation} \textbf{2014}, \emph{10}, 3151--3162\relax
\mciteBstWouldAddEndPuncttrue
\mciteSetBstMidEndSepPunct{\mcitedefaultmidpunct}
{\mcitedefaultendpunct}{\mcitedefaultseppunct}\relax
\EndOfBibitem
\bibitem[Fabiano \latin{et~al.}(2015)Fabiano, {Della Sala}, and
  Grabowski]{fabiano15}
Fabiano,~E.; {Della Sala},~F.; Grabowski,~I. Accurate non-covalent interaction
  energies via an efficient MP2 scaling procedure. \emph{Chemical Physics
  Letters} \textbf{2015}, \emph{635}, 262--267\relax
\mciteBstWouldAddEndPuncttrue
\mciteSetBstMidEndSepPunct{\mcitedefaultmidpunct}
{\mcitedefaultendpunct}{\mcitedefaultseppunct}\relax
\EndOfBibitem
\bibitem[Śmiga and Fabiano(2017)Śmiga, and Fabiano]{smiga17}
Śmiga,~S.; Fabiano,~E. Approximate solution of coupled cluster equations:
  application to the coupled cluster doubles method and non-covalent
  interacting systems. \emph{Phys. Chem. Chem. Phys.} \textbf{2017}, \emph{19},
  30249--30260\relax
\mciteBstWouldAddEndPuncttrue
\mciteSetBstMidEndSepPunct{\mcitedefaultmidpunct}
{\mcitedefaultendpunct}{\mcitedefaultseppunct}\relax
\EndOfBibitem
\bibitem[Fabiano and Cortona(2017)Fabiano, and Cortona]{fabiano17}
Fabiano,~E.; Cortona,~P. Dispersion corrections applied to the TCA family of
  exchange-correlation functionals. \emph{Theor Chem Acc} \textbf{2017},
  \emph{136}, 88\relax
\mciteBstWouldAddEndPuncttrue
\mciteSetBstMidEndSepPunct{\mcitedefaultmidpunct}
{\mcitedefaultendpunct}{\mcitedefaultseppunct}\relax
\EndOfBibitem
\bibitem[Riley \latin{et~al.}(2012)Riley, Platts, Řezáč, Hobza, and
  Hill]{riley12}
Riley,~K.~E.; Platts,~J.~A.; Řezáč,~J.; Hobza,~P.; Hill,~J.~G. Assessment of
  the Performance of MP2 and MP2 Variants for the Treatment of Noncovalent
  Interactions. \emph{The Journal of Physical Chemistry A} \textbf{2012},
  \emph{116}, 4159--4169\relax
\mciteBstWouldAddEndPuncttrue
\mciteSetBstMidEndSepPunct{\mcitedefaultmidpunct}
{\mcitedefaultendpunct}{\mcitedefaultseppunct}\relax
\EndOfBibitem
\bibitem[Hobza and Zahradnik(1988)Hobza, and Zahradnik]{hobza88}
Hobza,~P.; Zahradnik,~R. Intermolecular interactions between medium-sized
  systems. Nonempirical and empirical calculations of interaction energies.
  Successes and failures. \emph{Chemical Reviews} \textbf{1988}, \emph{88},
  871--897\relax
\mciteBstWouldAddEndPuncttrue
\mciteSetBstMidEndSepPunct{\mcitedefaultmidpunct}
{\mcitedefaultendpunct}{\mcitedefaultseppunct}\relax
\EndOfBibitem
\bibitem[Nguyen \latin{et~al.}(2020)Nguyen, Chen, Agee, Burow, Tang, and
  Furche]{Ngu-Fur-JCTC-2020}
Nguyen,~B.~D.; Chen,~G.~P.; Agee,~M.~M.; Burow,~A.~M.; Tang,~M.~P.; Furche,~F.
  Divergence of Many-Body Perturbation Theory for Noncovalent Interactions of
  Large Molecules. \emph{Journal of Chemical Theory and Computation}
  \textbf{2020}, \emph{16}, 2258--2273\relax
\mciteBstWouldAddEndPuncttrue
\mciteSetBstMidEndSepPunct{\mcitedefaultmidpunct}
{\mcitedefaultendpunct}{\mcitedefaultseppunct}\relax
\EndOfBibitem
\bibitem[Vuckovic \latin{et~al.}(2018)Vuckovic, Gori-Giorgi, Della~Sala, and
  Fabiano]{VucGorDelFab-JPCL-18}
Vuckovic,~S.; Gori-Giorgi,~P.; Della~Sala,~F.; Fabiano,~E. Restoring size
  consistency of approximate functionals constructed from the adiabatic
  connection. \emph{J. Phys. Chem. Lett.} \textbf{2018}, \emph{9},
  3137--3142\relax
\mciteBstWouldAddEndPuncttrue
\mciteSetBstMidEndSepPunct{\mcitedefaultmidpunct}
{\mcitedefaultendpunct}{\mcitedefaultseppunct}\relax
\EndOfBibitem
\bibitem[Grimme(2006)]{Gri-JCP-2006}
Grimme,~S. Semiempirical hybrid density functional with perturbative
  second-order correlation. \emph{The Journal of Chemical Physics}
  \textbf{2006}, \emph{124}, 034108\relax
\mciteBstWouldAddEndPuncttrue
\mciteSetBstMidEndSepPunct{\mcitedefaultmidpunct}
{\mcitedefaultendpunct}{\mcitedefaultseppunct}\relax
\EndOfBibitem
\bibitem[Schwabe and Grimme(2007)Schwabe, and Grimme]{SchGri-RSC-2007}
Schwabe,~T.; Grimme,~S. Double-hybrid density functionals with long-range
  dispersion corrections: higher accuracy and extended applicability.
  \emph{Phys. Chem. Chem. Phys.} \textbf{2007}, \emph{9}, 3397--3406\relax
\mciteBstWouldAddEndPuncttrue
\mciteSetBstMidEndSepPunct{\mcitedefaultmidpunct}
{\mcitedefaultendpunct}{\mcitedefaultseppunct}\relax
\EndOfBibitem
\bibitem[Caldeweyher \latin{et~al.}(2019)Caldeweyher, Ehlert, Hansen,
  Neugebauer, Spicher, Bannwarth, and Grimme]{CalEhlHanNeuSpiBanGri-JCP-2019}
Caldeweyher,~E.; Ehlert,~S.; Hansen,~A.; Neugebauer,~H.; Spicher,~S.;
  Bannwarth,~C.; Grimme,~S. A generally applicable atomic-charge dependent
  London dispersion correction. \emph{The Journal of Chemical Physics}
  \textbf{2019}, \emph{150}, 154122\relax
\mciteBstWouldAddEndPuncttrue
\mciteSetBstMidEndSepPunct{\mcitedefaultmidpunct}
{\mcitedefaultendpunct}{\mcitedefaultseppunct}\relax
\EndOfBibitem
\bibitem[Caldeweyher \latin{et~al.}(2020)Caldeweyher, Mewes, Ehlert, and
  Grimme]{CalMewEhlGri-PCCP-2020}
Caldeweyher,~E.; Mewes,~J.-M.; Ehlert,~S.; Grimme,~S. Extension and evaluation
  of the D4 London-dispersion model for periodic systems. \emph{Phys. Chem.
  Chem. Phys.} \textbf{2020}, \emph{22}, 8499--8512\relax
\mciteBstWouldAddEndPuncttrue
\mciteSetBstMidEndSepPunct{\mcitedefaultmidpunct}
{\mcitedefaultendpunct}{\mcitedefaultseppunct}\relax
\EndOfBibitem
\bibitem[Zhang \latin{et~al.}(2009)Zhang, Xu, and Goddard]{XYG3}
Zhang,~Y.; Xu,~X.; Goddard,~W.~A. Doubly hybrid density functional for accurate
  descriptions of nonbond interactions, thermochemistry, and thermochemical
  kinetics. \emph{Proceedings of the National Academy of Sciences}
  \textbf{2009}, \emph{106}, 4963--4968\relax
\mciteBstWouldAddEndPuncttrue
\mciteSetBstMidEndSepPunct{\mcitedefaultmidpunct}
{\mcitedefaultendpunct}{\mcitedefaultseppunct}\relax
\EndOfBibitem
\bibitem[Zhang and Xu(2021)Zhang, and Xu]{XYG7}
Zhang,~I.~Y.; Xu,~X. Exploring the Limits of the {XYG}3-Type Doubly Hybrid
  Approximations for the Main-Group Chemistry: The {xDH}@B3LYP Model. \emph{The
  Journal of Physical Chemistry Letters} \textbf{2021}, \emph{12},
  2638--2644\relax
\mciteBstWouldAddEndPuncttrue
\mciteSetBstMidEndSepPunct{\mcitedefaultmidpunct}
{\mcitedefaultendpunct}{\mcitedefaultseppunct}\relax
\EndOfBibitem
\bibitem[Santra \latin{et~al.}(2019)Santra, Sylvetsky, and Martin]{SSM19}
Santra,~G.; Sylvetsky,~N.; Martin,~J. M.~L. Minimally Empirical Double-Hybrid
  Functionals Trained against the {GMTKN}55 Database: {revDSD}-{PBEP}86-D4,
  {revDOD}-{PBE}-D4, and {DOD}-{SCAN}-D4. \emph{The Journal of Physical
  Chemistry A} \textbf{2019}, \emph{123}, 5129--5143\relax
\mciteBstWouldAddEndPuncttrue
\mciteSetBstMidEndSepPunct{\mcitedefaultmidpunct}
{\mcitedefaultendpunct}{\mcitedefaultseppunct}\relax
\EndOfBibitem
\bibitem[Seidl \latin{et~al.}(2018)Seidl, Giarrusso, Vuckovic, Fabiano, and
  Gori-Giorgi]{SeiGiaVucFabGor-JCP-18}
Seidl,~M.; Giarrusso,~S.; Vuckovic,~S.; Fabiano,~E.; Gori-Giorgi,~P.
  Communication: Strong-interaction limit of an adiabatic connection in
  Hartree-Fock theory. \emph{The Journal of Chemical Physics} \textbf{2018},
  \emph{149}, 241101\relax
\mciteBstWouldAddEndPuncttrue
\mciteSetBstMidEndSepPunct{\mcitedefaultmidpunct}
{\mcitedefaultendpunct}{\mcitedefaultseppunct}\relax
\EndOfBibitem
\bibitem[Daas \latin{et~al.}(2020)Daas, Grossi, Vuckovic, Musslimani, Kooi,
  Seidl, Giesbertz, and Gori-Giorgi]{DaaGroVucMusKooSeiGieGor-JCP-20}
Daas,~T.~J.; Grossi,~J.; Vuckovic,~S.; Musslimani,~Z.~H.; Kooi,~D.~P.;
  Seidl,~M.; Giesbertz,~K. J.~H.; Gori-Giorgi,~P. Large coupling-strength
  expansion of the M{\o}ller{\textendash}Plesset adiabatic connection: From
  paradigmatic cases to variational expressions for the leading terms.
  \emph{The Journal of Chemical Physics} \textbf{2020}, \emph{153},
  214112\relax
\mciteBstWouldAddEndPuncttrue
\mciteSetBstMidEndSepPunct{\mcitedefaultmidpunct}
{\mcitedefaultendpunct}{\mcitedefaultseppunct}\relax
\EndOfBibitem
\bibitem[Seidl \latin{et~al.}(1999)Seidl, Perdew, and Levy]{SeiPerLev-PRA-99}
Seidl,~M.; Perdew,~J.~P.; Levy,~M. Strictly correlated electrons in
  density-functional theory. \emph{Phys. Rev. A} \textbf{1999}, \emph{59},
  51--54\relax
\mciteBstWouldAddEndPuncttrue
\mciteSetBstMidEndSepPunct{\mcitedefaultmidpunct}
{\mcitedefaultendpunct}{\mcitedefaultseppunct}\relax
\EndOfBibitem
\bibitem[Sedlak \latin{et~al.}(2013)Sedlak, Janowski, Pito{\v{n}}{\'{a}}k,
  {\v{R}}ez{\'{a}}{\v{c}}, Pulay, and Hobza]{SedJanPitRezPulHob-ACS-2013}
Sedlak,~R.; Janowski,~T.; Pito{\v{n}}{\'{a}}k,~M.; {\v{R}}ez{\'{a}}{\v{c}},~J.;
  Pulay,~P.; Hobza,~P. Accuracy of Quantum Chemical Methods for Large
  Noncovalent Complexes. \emph{Journal of Chemical Theory and Computation}
  \textbf{2013}, \emph{9}, 3364--3374\relax
\mciteBstWouldAddEndPuncttrue
\mciteSetBstMidEndSepPunct{\mcitedefaultmidpunct}
{\mcitedefaultendpunct}{\mcitedefaultseppunct}\relax
\EndOfBibitem
\bibitem[Al-Hamdani \latin{et~al.}(2020)Al-Hamdani, Nagy, Barton, Kállay,
  Brandenburg, and Tkatchenko]{HamNagBarKalBraTka-unknown-2020}
Al-Hamdani,~Y.~S.; Nagy,~P.~R.; Barton,~D.; Kállay,~M.; Brandenburg,~J.~G.;
  Tkatchenko,~A. Interactions between Large Molecules: Puzzle for Reference
  Quantum-Mechanical Methods. 2020\relax
\mciteBstWouldAddEndPuncttrue
\mciteSetBstMidEndSepPunct{\mcitedefaultmidpunct}
{\mcitedefaultendpunct}{\mcitedefaultseppunct}\relax
\EndOfBibitem
\bibitem[Grimme \latin{et~al.}(2015)Grimme, Brandenburg, Bannwarth, and
  Hansen]{GriBraBanHan-JCP-2015}
Grimme,~S.; Brandenburg,~J.~G.; Bannwarth,~C.; Hansen,~A. Consistent structures
  and interactions by density functional theory with small atomic orbital basis
  sets. \emph{The Journal of Chemical Physics} \textbf{2015}, \emph{143},
  054107\relax
\mciteBstWouldAddEndPuncttrue
\mciteSetBstMidEndSepPunct{\mcitedefaultmidpunct}
{\mcitedefaultendpunct}{\mcitedefaultseppunct}\relax
\EndOfBibitem
\bibitem[Tkatchenko and Scheffler(2009)Tkatchenko, and
  Scheffler]{TkaSch-PRL-2009}
Tkatchenko,~A.; Scheffler,~M. Accurate Molecular Van Der Waals Interactions
  from Ground-State Electron Density and Free-Atom Reference Data. \emph{Phys.
  Rev. Lett.} \textbf{2009}, \emph{102}, 073005\relax
\mciteBstWouldAddEndPuncttrue
\mciteSetBstMidEndSepPunct{\mcitedefaultmidpunct}
{\mcitedefaultendpunct}{\mcitedefaultseppunct}\relax
\EndOfBibitem
\bibitem[Pernal(2018)]{Per-IJQC-18}
Pernal,~K. Correlation energy from random phase approximations: A reduced
  density matrices perspective. \emph{Int. J. Quantum. Chem.} \textbf{2018},
  \emph{118}, e25462\relax
\mciteBstWouldAddEndPuncttrue
\mciteSetBstMidEndSepPunct{\mcitedefaultmidpunct}
{\mcitedefaultendpunct}{\mcitedefaultseppunct}\relax
\EndOfBibitem
\bibitem[Marie \latin{et~al.}(2021)Marie, Burton, and
  Loos]{MarBurLoo-JPCM-2021}
Marie,~A.; Burton,~H. G.~A.; Loos,~P.-F. Perturbation Theory in the Complex
  Plane: Exceptional Points and Where to Find Them. \emph{Journal of Physics:
  Condensed Matter} \textbf{2021}, \relax
\mciteBstWouldAddEndPunctfalse
\mciteSetBstMidEndSepPunct{\mcitedefaultmidpunct}
{}{\mcitedefaultseppunct}\relax
\EndOfBibitem
\bibitem[M{\"o}ller and Plesset(1934)M{\"o}ller, and Plesset]{Moller1934}
M{\"o}ller,~C.; Plesset,~M.~S. Note on an Approximation Treatment for
  Many-Electron Systems. \emph{Phys. Rev.} \textbf{1934}, \emph{46},
  618--622\relax
\mciteBstWouldAddEndPuncttrue
\mciteSetBstMidEndSepPunct{\mcitedefaultmidpunct}
{\mcitedefaultendpunct}{\mcitedefaultseppunct}\relax
\EndOfBibitem
\bibitem[Seidl \latin{et~al.}(2007)Seidl, Gori-Giorgi, and
  Savin]{SeiGorSav-PRA-07}
Seidl,~M.; Gori-Giorgi,~P.; Savin,~A. Strictly correlated electrons in
  density-functional theory: A general formulation with applications to
  spherical densities. \emph{Phys. Rev. A} \textbf{2007}, \emph{75},
  042511/12\relax
\mciteBstWouldAddEndPuncttrue
\mciteSetBstMidEndSepPunct{\mcitedefaultmidpunct}
{\mcitedefaultendpunct}{\mcitedefaultseppunct}\relax
\EndOfBibitem
\bibitem[Gori-Giorgi \latin{et~al.}(2009)Gori-Giorgi, Vignale, and
  Seidl]{GorVigSei-JCTC-09}
Gori-Giorgi,~P.; Vignale,~G.; Seidl,~M. Electronic Zero-Point Oscillations in
  the Strong-Interaction Limit of Density Functional Theory. \emph{J. Chem.
  Theory Comput.} \textbf{2009}, \emph{5}, 743--753\relax
\mciteBstWouldAddEndPuncttrue
\mciteSetBstMidEndSepPunct{\mcitedefaultmidpunct}
{\mcitedefaultendpunct}{\mcitedefaultseppunct}\relax
\EndOfBibitem
\bibitem[Langreth and Perdew(1975)Langreth, and Perdew]{LanPer-SSC-75}
Langreth,~D.~C.; Perdew,~J.~P. The exchange-correlation energy of a metallic
  surface. \emph{Solid. State Commun.} \textbf{1975}, \emph{17},
  1425--1429\relax
\mciteBstWouldAddEndPuncttrue
\mciteSetBstMidEndSepPunct{\mcitedefaultmidpunct}
{\mcitedefaultendpunct}{\mcitedefaultseppunct}\relax
\EndOfBibitem
\bibitem[Gunnarsson and Lundqvist(1976)Gunnarsson, and
  Lundqvist]{GunLun-PRB-76}
Gunnarsson,~O.; Lundqvist,~B.~I. Exchange and correlation in atoms, molecules,
  and solids by the spin-density-functional formalism. \emph{Phys. Rev. B}
  \textbf{1976}, \emph{13}, 4274--4298\relax
\mciteBstWouldAddEndPuncttrue
\mciteSetBstMidEndSepPunct{\mcitedefaultmidpunct}
{\mcitedefaultendpunct}{\mcitedefaultseppunct}\relax
\EndOfBibitem
\bibitem[Gori-Giorgi \latin{et~al.}(2008)Gori-Giorgi, Seidl, and
  Savin]{GorSeiSav-PCCP-08}
Gori-Giorgi,~P.; Seidl,~M.; Savin,~A. \emph{Phys. Chem. Chem. Phys.}
  \textbf{2008}, \emph{{10}}, 3440\relax
\mciteBstWouldAddEndPuncttrue
\mciteSetBstMidEndSepPunct{\mcitedefaultmidpunct}
{\mcitedefaultendpunct}{\mcitedefaultseppunct}\relax
\EndOfBibitem
\bibitem[Gori-Giorgi \latin{et~al.}(2009)Gori-Giorgi, Seidl, and
  Vignale]{GorSeiVig-PRL-09}
Gori-Giorgi,~P.; Seidl,~M.; Vignale,~G. Density-Functional Theory for Strongly
  Interacting Electrons. \emph{Phys. Rev. Lett.} \textbf{2009}, \emph{{103}},
  166402\relax
\mciteBstWouldAddEndPuncttrue
\mciteSetBstMidEndSepPunct{\mcitedefaultmidpunct}
{\mcitedefaultendpunct}{\mcitedefaultseppunct}\relax
\EndOfBibitem
\bibitem[Grossi \latin{et~al.}(2017)Grossi, Kooi, Giesbertz, Seidl, Cohen,
  Mori-S{\'a}nchez, and Gori-Giorgi]{GroKooGieSeiCohMorGor-JCTC-17}
Grossi,~J.; Kooi,~D.~P.; Giesbertz,~K. J.~H.; Seidl,~M.; Cohen,~A.~J.;
  Mori-S{\'a}nchez,~P.; Gori-Giorgi,~P. Fermionic statistics in the strongly
  correlated limit of Density Functional Theory. \emph{J. Chem. Theory Comput.}
  \textbf{2017}, \emph{13}, 6089--6100\relax
\mciteBstWouldAddEndPuncttrue
\mciteSetBstMidEndSepPunct{\mcitedefaultmidpunct}
{\mcitedefaultendpunct}{\mcitedefaultseppunct}\relax
\EndOfBibitem
\bibitem[G\"{o}rling and Levy(1993)G\"{o}rling, and Levy]{GorLev-PRB-93}
G\"{o}rling,~A.; Levy,~M. \emph{Phys. Rev. B} \textbf{1993}, \emph{47},
  13105\relax
\mciteBstWouldAddEndPuncttrue
\mciteSetBstMidEndSepPunct{\mcitedefaultmidpunct}
{\mcitedefaultendpunct}{\mcitedefaultseppunct}\relax
\EndOfBibitem
\bibitem[G\"{o}rling and Levy(1994)G\"{o}rling, and Levy]{GorLev-PRA-94}
G\"{o}rling,~A.; Levy,~M. Exact Kohn-Sham scheme based on perturbation theory.
  \emph{Phys. Rev. A} \textbf{1994}, \emph{50}, 196\relax
\mciteBstWouldAddEndPuncttrue
\mciteSetBstMidEndSepPunct{\mcitedefaultmidpunct}
{\mcitedefaultendpunct}{\mcitedefaultseppunct}\relax
\EndOfBibitem
\bibitem[Seidl \latin{et~al.}(2000)Seidl, Perdew, and Kurth]{SeiPerKur-PRA-00}
Seidl,~M.; Perdew,~J.~P.; Kurth,~S. \emph{Phys. Rev. A} \textbf{2000},
  \emph{{62}}, 012502\relax
\mciteBstWouldAddEndPuncttrue
\mciteSetBstMidEndSepPunct{\mcitedefaultmidpunct}
{\mcitedefaultendpunct}{\mcitedefaultseppunct}\relax
\EndOfBibitem
\bibitem[Wagner and Gori-Giorgi(2014)Wagner, and Gori-Giorgi]{WagGor-PRA-14}
Wagner,~L.~O.; Gori-Giorgi,~P. Electron avoidance: A nonlocal radius for strong
  correlation. \emph{Phys. Rev. A} \textbf{2014}, \emph{90}, 052512\relax
\mciteBstWouldAddEndPuncttrue
\mciteSetBstMidEndSepPunct{\mcitedefaultmidpunct}
{\mcitedefaultendpunct}{\mcitedefaultseppunct}\relax
\EndOfBibitem
\bibitem[Bahmann \latin{et~al.}(2016)Bahmann, Zhou, and
  Ernzerhof]{BahZhoErn-JCP-16}
Bahmann,~H.; Zhou,~Y.; Ernzerhof,~M. The shell model for the
  exchange-correlation hole in the strong-correlation limit. \emph{J. Chem.
  Phys.} \textbf{2016}, \emph{145}, 124104\relax
\mciteBstWouldAddEndPuncttrue
\mciteSetBstMidEndSepPunct{\mcitedefaultmidpunct}
{\mcitedefaultendpunct}{\mcitedefaultseppunct}\relax
\EndOfBibitem
\bibitem[Vuckovic and Gori-Giorgi(2017)Vuckovic, and
  Gori-Giorgi]{VucGor-JPCL-17}
Vuckovic,~S.; Gori-Giorgi,~P. Simple Fully Nonlocal Density Functionals for
  Electronic Repulsion Energy. \emph{The Journal of Physical Chemistry Letters}
  \textbf{2017}, \emph{8}, 2799--2805\relax
\mciteBstWouldAddEndPuncttrue
\mciteSetBstMidEndSepPunct{\mcitedefaultmidpunct}
{\mcitedefaultendpunct}{\mcitedefaultseppunct}\relax
\EndOfBibitem
\bibitem[Gould and Vuckovic(2019)Gould, and Vuckovic]{GouVuc-JCP-2019}
Gould,~T.; Vuckovic,~S. Range-separation and the multiple radii functional
  approximation inspired by the strongly interacting limit of density
  functional theory. \emph{The Journal of Chemical Physics} \textbf{2019},
  \emph{151}, 184101\relax
\mciteBstWouldAddEndPuncttrue
\mciteSetBstMidEndSepPunct{\mcitedefaultmidpunct}
{\mcitedefaultendpunct}{\mcitedefaultseppunct}\relax
\EndOfBibitem
\bibitem[Becke(1993)]{Bec-JCP-93a}
Becke,~A.~D. A new mixing of Hartree--Fock and local density-functional
  theories. \emph{J. Chem. Phys.} \textbf{1993}, \emph{98}, 1372\relax
\mciteBstWouldAddEndPuncttrue
\mciteSetBstMidEndSepPunct{\mcitedefaultmidpunct}
{\mcitedefaultendpunct}{\mcitedefaultseppunct}\relax
\EndOfBibitem
\bibitem[Becke(1993)]{Bec-JCP-93}
Becke,~A.~D. Density-functional thermochemistry. III. The role of exact
  exchange. \emph{J. Chem. Phys.} \textbf{1993}, \emph{98}, 5648\relax
\mciteBstWouldAddEndPuncttrue
\mciteSetBstMidEndSepPunct{\mcitedefaultmidpunct}
{\mcitedefaultendpunct}{\mcitedefaultseppunct}\relax
\EndOfBibitem
\bibitem[Perdew \latin{et~al.}(1996)Perdew, Ernzerhof, and
  Burke]{PerErnBur-JCP-96}
Perdew,~J.~P.; Ernzerhof,~M.; Burke,~K. Rationale for mixing exact exchange
  with density functional approximations. \emph{J. Chem. Phys.} \textbf{1996},
  \emph{105}, 9982--9985\relax
\mciteBstWouldAddEndPuncttrue
\mciteSetBstMidEndSepPunct{\mcitedefaultmidpunct}
{\mcitedefaultendpunct}{\mcitedefaultseppunct}\relax
\EndOfBibitem
\bibitem[Sharkas \latin{et~al.}(2011)Sharkas, Toulouse, and
  Savin]{ShaTouSav-JCP-11}
Sharkas,~K.; Toulouse,~J.; Savin,~A. \emph{J. Chem. Phys.} \textbf{2011},
  \emph{{134}}, 064113\relax
\mciteBstWouldAddEndPuncttrue
\mciteSetBstMidEndSepPunct{\mcitedefaultmidpunct}
{\mcitedefaultendpunct}{\mcitedefaultseppunct}\relax
\EndOfBibitem
\bibitem[Goerigk and Grimme(2010)Goerigk, and Grimme]{LarGri-JCTC-10}
Goerigk,~L.; Grimme,~S. Efficient and Accurate Double-Hybrid-Meta-GGA Density
  Functionals Evaluation with the Extended GMTKN30 Database for General Main
  Group Thermochemistry, Kinetics, and Noncovalent Interactions. \emph{J. Chem.
  Theory Comput.} \textbf{2010}, \emph{7}, 291--309\relax
\mciteBstWouldAddEndPuncttrue
\mciteSetBstMidEndSepPunct{\mcitedefaultmidpunct}
{\mcitedefaultendpunct}{\mcitedefaultseppunct}\relax
\EndOfBibitem
\bibitem[Su and Xu(2014)Su, and Xu]{SuXu-JCP-14}
Su,~N.~Q.; Xu,~X. Construction of a parameter-free doubly hybrid density
  functional from adiabatic connection. \emph{J. Chem. Phys.} \textbf{2014},
  \emph{140}, 18A512\relax
\mciteBstWouldAddEndPuncttrue
\mciteSetBstMidEndSepPunct{\mcitedefaultmidpunct}
{\mcitedefaultendpunct}{\mcitedefaultseppunct}\relax
\EndOfBibitem
\bibitem[Vuckovic \latin{et~al.}(2016)Vuckovic, Irons, Savin, Teale, and
  Gori-Giorgi]{VucIroSavTeaGor-JCTC-16}
Vuckovic,~S.; Irons,~T. J.~P.; Savin,~A.; Teale,~A.~M.; Gori-Giorgi,~P.
  Exchange--correlation functionals via local interpolation along the adiabatic
  connection. \emph{J. Chem. Theory Comput.} \textbf{2016}, \emph{12},
  2598--2610\relax
\mciteBstWouldAddEndPuncttrue
\mciteSetBstMidEndSepPunct{\mcitedefaultmidpunct}
{\mcitedefaultendpunct}{\mcitedefaultseppunct}\relax
\EndOfBibitem
\bibitem[Pastorczak \latin{et~al.}(2019)Pastorczak, Hapka, Veis, and
  Pernal]{PasHapVeiPer-JPCL-2019}
Pastorczak,~E.; Hapka,~M.; Veis,~L.; Pernal,~K. Capturing the Dynamic
  Correlation for Arbitrary Spin-Symmetry CASSCF Reference with Adiabatic
  Connection Approaches: Insights into the Electronic Structure of the
  Tetramethyleneethane Diradical. \emph{The Journal of Physical Chemistry
  Letters} \textbf{2019}, \emph{10}, 4668--4674\relax
\mciteBstWouldAddEndPuncttrue
\mciteSetBstMidEndSepPunct{\mcitedefaultmidpunct}
{\mcitedefaultendpunct}{\mcitedefaultseppunct}\relax
\EndOfBibitem
\bibitem[Maradzike \latin{et~al.}(2020)Maradzike, Hapka, Pernal, and
  DePrince]{MarHapPerDeP-JCTC-2020}
Maradzike,~E.; Hapka,~M.; Pernal,~K.; DePrince,~A.~E. Reduced Density
  Matrix-Driven Complete Active Apace Self-Consistent Field Corrected for
  Dynamic Correlation from the Adiabatic Connection. \emph{Journal of Chemical
  Theory and Computation} \textbf{2020}, \emph{16}, 4351--4360\relax
\mciteBstWouldAddEndPuncttrue
\mciteSetBstMidEndSepPunct{\mcitedefaultmidpunct}
{\mcitedefaultendpunct}{\mcitedefaultseppunct}\relax
\EndOfBibitem
\bibitem[Seidl \latin{et~al.}(2000)Seidl, Perdew, and Kurth]{SeiPerKur-PRL-00}
Seidl,~M.; Perdew,~J.~P.; Kurth,~S. Simulation of All-Order Density-Functional
  Perturbation Theory, Using the Second Order and the Strong-Correlation Limit.
  \emph{Phys. Rev. Lett.} \textbf{2000}, \emph{84}, 5070--5073\relax
\mciteBstWouldAddEndPuncttrue
\mciteSetBstMidEndSepPunct{\mcitedefaultmidpunct}
{\mcitedefaultendpunct}{\mcitedefaultseppunct}\relax
\EndOfBibitem
\bibitem[Vuckovic \latin{et~al.}(2017)Vuckovic, Irons, Wagner, Teale, and
  Gori-Giorgi]{VucIroWagTeaGor-PCCP-17}
Vuckovic,~S.; Irons,~T. J.~P.; Wagner,~L.~O.; Teale,~A.~M.; Gori-Giorgi,~P.
  Interpolated energy densities{,} correlation indicators and lower bounds from
  approximations to the strong coupling limit of DFT. \emph{Phys. Chem. Chem.
  Phys.} \textbf{2017}, \emph{19}, 6169--6183\relax
\mciteBstWouldAddEndPuncttrue
\mciteSetBstMidEndSepPunct{\mcitedefaultmidpunct}
{\mcitedefaultendpunct}{\mcitedefaultseppunct}\relax
\EndOfBibitem
\bibitem[Vuckovic \latin{et~al.}(2020)Vuckovic, Fabiano, Gori-Giorgi, and
  Burke]{VucFabGorBur-JCTC-20}
Vuckovic,~S.; Fabiano,~E.; Gori-Giorgi,~P.; Burke,~K. MAP: An MP2 Accuracy
  Predictor for Weak Interactions from Adiabatic Connection Theory.
  \emph{Journal of Chemical Theory and Computation} \textbf{2020}, \emph{16},
  4141--4149\relax
\mciteBstWouldAddEndPuncttrue
\mciteSetBstMidEndSepPunct{\mcitedefaultmidpunct}
{\mcitedefaultendpunct}{\mcitedefaultseppunct}\relax
\EndOfBibitem
\bibitem[Constantin(2019)]{Con-PRB-2019}
Constantin,~L.~A. Correlation energy functionals from adiabatic connection
  formalism. \emph{Phys. Rev. B} \textbf{2019}, \emph{99}, 085117\relax
\mciteBstWouldAddEndPuncttrue
\mciteSetBstMidEndSepPunct{\mcitedefaultmidpunct}
{\mcitedefaultendpunct}{\mcitedefaultseppunct}\relax
\EndOfBibitem
\bibitem[Mirtschink \latin{et~al.}(2012)Mirtschink, Seidl, and
  Gori-Giorgi]{MirSeiGor-JCTC-12}
Mirtschink,~A.; Seidl,~M.; Gori-Giorgi,~P. Energy densities in the
  strong-interaction limit of density functional theory. \emph{J. Chem. Theory
  Comput.} \textbf{2012}, \emph{8}, 3097--3107\relax
\mciteBstWouldAddEndPuncttrue
\mciteSetBstMidEndSepPunct{\mcitedefaultmidpunct}
{\mcitedefaultendpunct}{\mcitedefaultseppunct}\relax
\EndOfBibitem
\bibitem[Řezáč \latin{et~al.}(2011)Řezáč, Riley, and
  Hobza]{RezRilHob-JCTC-2011}
Řezáč,~J.; Riley,~K.~E.; Hobza,~P. S66: A Well-balanced Database of
  Benchmark Interaction Energies Relevant to Biomolecular Structures.
  \emph{Journal of Chemical Theory and Computation} \textbf{2011}, \emph{7},
  2427--2438\relax
\mciteBstWouldAddEndPuncttrue
\mciteSetBstMidEndSepPunct{\mcitedefaultmidpunct}
{\mcitedefaultendpunct}{\mcitedefaultseppunct}\relax
\EndOfBibitem
\bibitem[Jure{\v{c}}ka \latin{et~al.}(2006)Jure{\v{c}}ka, {\v{S}}poner,
  {\v{C}}ern{\'{y}}, and Hobza]{S22}
Jure{\v{c}}ka,~P.; {\v{S}}poner,~J.; {\v{C}}ern{\'{y}},~J.; Hobza,~P. Benchmark
  database of accurate ({MP}2 and {CCSD}(T) complete basis set limit)
  interaction energies of small model complexes, {DNA} base pairs, and amino
  acid pairs. \emph{Phys. Chem. Chem. Phys.} \textbf{2006}, \emph{8},
  1985--1993\relax
\mciteBstWouldAddEndPuncttrue
\mciteSetBstMidEndSepPunct{\mcitedefaultmidpunct}
{\mcitedefaultendpunct}{\mcitedefaultseppunct}\relax
\EndOfBibitem
\bibitem[Takatani \latin{et~al.}(2010)Takatani, Hohenstein, Malagoli, Marshall,
  and Sherrill]{S22ref}
Takatani,~T.; Hohenstein,~E.~G.; Malagoli,~M.; Marshall,~M.~S.; Sherrill,~C.~D.
  Basis set consistent revision of the S22 test set of noncovalent interaction
  energies. \emph{The Journal of chemical physics} \textbf{2010}, \emph{132},
  144104\relax
\mciteBstWouldAddEndPuncttrue
\mciteSetBstMidEndSepPunct{\mcitedefaultmidpunct}
{\mcitedefaultendpunct}{\mcitedefaultseppunct}\relax
\EndOfBibitem
\bibitem[sup()]{sup}
See Supplemental information at .... for further information, figures and
  tables.\relax
\mciteBstWouldAddEndPunctfalse
\mciteSetBstMidEndSepPunct{\mcitedefaultmidpunct}
{}{\mcitedefaultseppunct}\relax
\EndOfBibitem
\bibitem[Vuckovic and Burke(2020)Vuckovic, and Burke]{VucBur-JPCL-20}
Vuckovic,~S.; Burke,~K. Quantifying and Understanding Errors in Molecular
  Geometries. \emph{The Journal of Physical Chemistry Letters} \textbf{2020},
  \emph{11}, 9957--9964\relax
\mciteBstWouldAddEndPuncttrue
\mciteSetBstMidEndSepPunct{\mcitedefaultmidpunct}
{\mcitedefaultendpunct}{\mcitedefaultseppunct}\relax
\EndOfBibitem
\bibitem[Zhao and Truhlar(2005)Zhao, and Truhlar]{ZhaTru-JCTC-2005}
Zhao,~Y.; Truhlar,~D.~G. Benchmark Databases for Nonbonded Interactions and
  Their Use To Test Density Functional Theory. \emph{Journal of Chemical Theory
  and Computation} \textbf{2005}, \emph{1}, 415--432\relax
\mciteBstWouldAddEndPuncttrue
\mciteSetBstMidEndSepPunct{\mcitedefaultmidpunct}
{\mcitedefaultendpunct}{\mcitedefaultseppunct}\relax
\EndOfBibitem
\bibitem[Zhao \latin{et~al.}(2005)Zhao, Schultz, and
  Truhlar]{ZhaSchTru-JCP-2005}
Zhao,~Y.; Schultz,~N.~E.; Truhlar,~D.~G. Exchange-correlation functional with
  broad accuracy for metallic and nonmetallic compounds, kinetics, and
  noncovalent interactions. \emph{The Journal of Chemical Physics}
  \textbf{2005}, \emph{123}, 161103\relax
\mciteBstWouldAddEndPuncttrue
\mciteSetBstMidEndSepPunct{\mcitedefaultmidpunct}
{\mcitedefaultendpunct}{\mcitedefaultseppunct}\relax
\EndOfBibitem
\bibitem[Zhao \latin{et~al.}(2006)Zhao, Schultz, and
  Truhlar]{ZhaSchTru-JCTC-06}
Zhao,~Y.; Schultz,~N.~E.; Truhlar,~D.~G. \emph{J. Chem. Theory Comput.}
  \textbf{2006}, \emph{2}, 364\relax
\mciteBstWouldAddEndPuncttrue
\mciteSetBstMidEndSepPunct{\mcitedefaultmidpunct}
{\mcitedefaultendpunct}{\mcitedefaultseppunct}\relax
\EndOfBibitem
\bibitem[Zhang \latin{et~al.}(2011)Zhang, Xu, Jung, and
  Goddard]{ZhaXuJunGod-PNAS-2011}
Zhang,~I.~Y.; Xu,~X.; Jung,~Y.; Goddard,~W.~A. A fast doubly hybrid density
  functional method close to chemical accuracy using a local opposite spin
  ansatz. \emph{Proceedings of the National Academy of Sciences} \textbf{2011},
  \emph{108}, 19896--19900\relax
\mciteBstWouldAddEndPuncttrue
\mciteSetBstMidEndSepPunct{\mcitedefaultmidpunct}
{\mcitedefaultendpunct}{\mcitedefaultseppunct}\relax
\EndOfBibitem
\bibitem[Kim \latin{et~al.}(2018)Kim, Song, Sim, and Burke]{KSSB18}
Kim,~Y.; Song,~S.; Sim,~E.; Burke,~K. Halogen and Chalcogen Binding Dominated
  by Density-Driven Errors. \emph{The Journal of Physical Chemistry Letters}
  \textbf{2018}, \emph{10}, 295--301\relax
\mciteBstWouldAddEndPuncttrue
\mciteSetBstMidEndSepPunct{\mcitedefaultmidpunct}
{\mcitedefaultendpunct}{\mcitedefaultseppunct}\relax
\EndOfBibitem
\bibitem[Song \latin{et~al.}(2021)Song, Vuckovic, Sim, and
  Burke]{SonVucSimBur-JPCL-21}
Song,~S.; Vuckovic,~S.; Sim,~E.; Burke,~K. Density Sensitivity of Empirical
  Functionals. \emph{The Journal of Physical Chemistry Letters} \textbf{2021},
  \emph{12}, 800--807\relax
\mciteBstWouldAddEndPuncttrue
\mciteSetBstMidEndSepPunct{\mcitedefaultmidpunct}
{\mcitedefaultendpunct}{\mcitedefaultseppunct}\relax
\EndOfBibitem
\bibitem[Mehta \latin{et~al.}(2021)Mehta, Fellowes, WHITE, and
  Goerigk]{lars2021}
Mehta,~N.; Fellowes,~T.; WHITE,~J.; Goerigk,~L. TheCHAL336 Benchmark Set: How
  Well Do Quantum-Chemical Methods Describe Chalcogen-Bonding Interactions?
  \textbf{2021}, \relax
\mciteBstWouldAddEndPunctfalse
\mciteSetBstMidEndSepPunct{\mcitedefaultmidpunct}
{}{\mcitedefaultseppunct}\relax
\EndOfBibitem
\bibitem[Mardirossian and Head-Gordon(2016)Mardirossian, and Head-Gordon]{MH16}
Mardirossian,~N.; Head-Gordon,~M. $\omega$B97M-V: A combinatorially optimized,
  range-separated hybrid, meta-GGA density functional with VV10 nonlocal
  correlation. \emph{The Journal of Chemical Physics} \textbf{2016},
  \emph{144}, 214110\relax
\mciteBstWouldAddEndPuncttrue
\mciteSetBstMidEndSepPunct{\mcitedefaultmidpunct}
{\mcitedefaultendpunct}{\mcitedefaultseppunct}\relax
\EndOfBibitem
\bibitem[Mardirossian and Head-Gordon(2017)Mardirossian, and Head-Gordon]{MH17}
Mardirossian,~N.; Head-Gordon,~M. Thirty years of density functional theory in
  computational chemistry: an overview and extensive assessment of 200 density
  functionals. \emph{Molecular Physics} \textbf{2017}, \emph{115},
  2315--2372\relax
\mciteBstWouldAddEndPuncttrue
\mciteSetBstMidEndSepPunct{\mcitedefaultmidpunct}
{\mcitedefaultendpunct}{\mcitedefaultseppunct}\relax
\EndOfBibitem
\bibitem[Mardirossian and Head-Gordon(2018)Mardirossian, and
  Head-Gordon]{MHG18}
Mardirossian,~N.; Head-Gordon,~M. Survival of the most transferable at the top
  of Jacob's ladder: Defining and testing the $\upomega$B97M(2) double hybrid
  density functional. \emph{The Journal of Chemical Physics} \textbf{2018},
  \emph{148}, 241736\relax
\mciteBstWouldAddEndPuncttrue
\mciteSetBstMidEndSepPunct{\mcitedefaultmidpunct}
{\mcitedefaultendpunct}{\mcitedefaultseppunct}\relax
\EndOfBibitem
\bibitem[McGibbon \latin{et~al.}(2017)McGibbon, Taube, Donchev, Siva,
  Hern{\'{a}}ndez, Hargus, Law, Klepeis, and Shaw]{SNSMP2}
McGibbon,~R.~T.; Taube,~A.~G.; Donchev,~A.~G.; Siva,~K.; Hern{\'{a}}ndez,~F.;
  Hargus,~C.; Law,~K.-H.; Klepeis,~J.~L.; Shaw,~D.~E. Improving the accuracy of
  M{\o}ller-Plesset perturbation theory with neural networks. \emph{The Journal
  of Chemical Physics} \textbf{2017}, \emph{147}, 161725\relax
\mciteBstWouldAddEndPuncttrue
\mciteSetBstMidEndSepPunct{\mcitedefaultmidpunct}
{\mcitedefaultendpunct}{\mcitedefaultseppunct}\relax
\EndOfBibitem
\bibitem[Sch\"{u}tz \latin{et~al.}(1999)Sch\"{u}tz, Hetzer, and Werner]{SHW99}
Sch\"{u}tz,~M.; Hetzer,~G.; Werner,~H.-J. Low-order scaling local electron
  correlation methods. I. Linear scaling local {MP}2. \emph{The Journal of
  Chemical Physics} \textbf{1999}, \emph{111}, 5691--5705\relax
\mciteBstWouldAddEndPuncttrue
\mciteSetBstMidEndSepPunct{\mcitedefaultmidpunct}
{\mcitedefaultendpunct}{\mcitedefaultseppunct}\relax
\EndOfBibitem
\bibitem[Lee \latin{et~al.}(2000)Lee, Maslen, and Head-Gordon]{LMH00}
Lee,~M.~S.; Maslen,~P.~E.; Head-Gordon,~M. Closely approximating second-order
  Mo/ller{\textendash}Plesset perturbation theory with a local triatomics in
  molecules model. \emph{The Journal of Chemical Physics} \textbf{2000},
  \emph{112}, 3592--3601\relax
\mciteBstWouldAddEndPuncttrue
\mciteSetBstMidEndSepPunct{\mcitedefaultmidpunct}
{\mcitedefaultendpunct}{\mcitedefaultseppunct}\relax
\EndOfBibitem
\bibitem[Williams \latin{et~al.}(2020)Williams, Wiles, and
  Manby]{WilWilMan-JCTC-2020}
Williams,~Z.~M.; Wiles,~T.~C.; Manby,~F.~R. Accurate Hybrid Density Functionals
  with UW12 Correlation. \emph{Journal of Chemical Theory and Computation}
  \textbf{2020}, \emph{16}, 6176--6194\relax
\mciteBstWouldAddEndPuncttrue
\mciteSetBstMidEndSepPunct{\mcitedefaultmidpunct}
{\mcitedefaultendpunct}{\mcitedefaultseppunct}\relax
\EndOfBibitem
\bibitem[Furche \latin{et~al.}(2014)Furche, Ahlrichs, Hättig, Klopper, Sierka,
  and Weigend]{Furche2014}
Furche,~F.; Ahlrichs,~R.; Hättig,~C.; Klopper,~W.; Sierka,~M.; Weigend,~F.
  Turbomole. \emph{WIREs Computational Molecular Science} \textbf{2014},
  \emph{4}, 91--100\relax
\mciteBstWouldAddEndPuncttrue
\mciteSetBstMidEndSepPunct{\mcitedefaultmidpunct}
{\mcitedefaultendpunct}{\mcitedefaultseppunct}\relax
\EndOfBibitem
\bibitem[TUR()]{TURBOMOLE}
{TURBOMOLE V7.1 2010}, a development of {University of Karlsruhe} and
  {Forschungszentrum Karlsruhe GmbH}, 1989-2007, {TURBOMOLE GmbH}, since 2007;
  available from \\ {\tt http://www.turbomole.com}.\relax
\mciteBstWouldAddEndPunctfalse
\mciteSetBstMidEndSepPunct{\mcitedefaultmidpunct}
{}{\mcitedefaultseppunct}\relax
\EndOfBibitem
\bibitem[Fabiano \latin{et~al.}(2016)Fabiano, Gori-Giorgi, Seidl, and
  Della~Sala]{FabGorSeiDel-JCTC-16}
Fabiano,~E.; Gori-Giorgi,~P.; Seidl,~M.; Della~Sala,~F. Interaction-Strength
  Interpolation Method for Main-Group Chemistry: Benchmarking, Limitations, and
  Perspectives. \emph{J. Chem. Theory. Comput.} \textbf{2016}, \emph{12},
  4885--4896\relax
\mciteBstWouldAddEndPuncttrue
\mciteSetBstMidEndSepPunct{\mcitedefaultmidpunct}
{\mcitedefaultendpunct}{\mcitedefaultseppunct}\relax
\EndOfBibitem
\bibitem[Giarrusso \latin{et~al.}(2018)Giarrusso, Gori-Giorgi, Della~Sala, and
  Fabiano]{GiaGorDelFab-JCP-18}
Giarrusso,~S.; Gori-Giorgi,~P.; Della~Sala,~F.; Fabiano,~E. Assessment of
  interaction-strength interpolation formulas for gold and silver clusters.
  \emph{J. Chem. Phys.} \textbf{2018}, \emph{148}, 134106\relax
\mciteBstWouldAddEndPuncttrue
\mciteSetBstMidEndSepPunct{\mcitedefaultmidpunct}
{\mcitedefaultendpunct}{\mcitedefaultseppunct}\relax
\EndOfBibitem
\bibitem[Goerigk \latin{et~al.}(2017)Goerigk, Hansen, Bauer, Ehrlich, Najibi,
  and Grimme]{GoeHanBauEhrNajGri-RSC-2017}
Goerigk,~L.; Hansen,~A.; Bauer,~C.; Ehrlich,~S.; Najibi,~A.; Grimme,~S. A look
  at the density functional theory zoo with the advanced {GMTKN}55 database for
  general main group thermochemistry, kinetics and noncovalent interactions.
  \emph{Physical Chemistry Chemical Physics} \textbf{2017}, \emph{19},
  32184--32215\relax
\mciteBstWouldAddEndPuncttrue
\mciteSetBstMidEndSepPunct{\mcitedefaultmidpunct}
{\mcitedefaultendpunct}{\mcitedefaultseppunct}\relax
\EndOfBibitem
\bibitem[Calbo \latin{et~al.}(2015)Calbo, Ortí, Sancho-García, and
  Aragó]{CalOrtSanAra-JCTC-2015}
Calbo,~J.; Ortí,~E.; Sancho-García,~J.~C.; Aragó,~J. Accurate Treatment of
  Large Supramolecular Complexes by Double-Hybrid Density Functionals Coupled
  with Nonlocal van der Waals Corrections. \emph{Journal of Chemical Theory and
  Computation} \textbf{2015}, \emph{11}, 932--939\relax
\mciteBstWouldAddEndPuncttrue
\mciteSetBstMidEndSepPunct{\mcitedefaultmidpunct}
{\mcitedefaultendpunct}{\mcitedefaultseppunct}\relax
\EndOfBibitem
\bibitem[Kovács \latin{et~al.}(2017)Kovács, Cz.~Dobrowolski, Ostrowski, and
  Rode]{KovDobOstRod-2017-IJQC-2017}
Kovács,~A.; Cz.~Dobrowolski,~J.; Ostrowski,~S.; Rode,~J.~E. Benchmarking
  density functionals in conjunction with Grimme's dispersion correction for
  noble gas dimers (Ne2, Ar2, Kr2, Xe2, Rn2). \emph{International Journal of
  Quantum Chemistry} \textbf{2017}, \emph{117}, e25358\relax
\mciteBstWouldAddEndPuncttrue
\mciteSetBstMidEndSepPunct{\mcitedefaultmidpunct}
{\mcitedefaultendpunct}{\mcitedefaultseppunct}\relax
\EndOfBibitem
\end{mcitethebibliography}

\end{document}


\begin{table}
\caption{MAEs in kcal/mol of MP2, SPL, MPACF-1 and SPL2 for the three subsets of the S66 dataset.}
\begin{tabular}{c|ccc}
 \text{method} & \text{H-bonds} & \text{dispersion} & \text{others} \\
 \hline
 \text{MP2} & 0.18 & 0.82 & 0.41 \\
 \text{SPL} & 0.42 & 0.42 & 0.19 \\
 \text{MPACF-1} & 0.15 & 0.45 & 0.17 \\
 \text{SPL2} & 0.19 & 0.30 & 0.12 \\
\end{tabular}
\end{table}

\begin{table}[]
\caption{MAEs of 3 different reference calculations relative to each other, as well as the MAEs of MP2, SPL, SPL2, MPACF-1 relative to each of the three references for the L7 dataset. Regardless of the reference, SPL gives improvement over MP2 and SPL2 and MPACF-1 give improvements over SPL. Data from ref. 1 for L7 were obtained from DLPNO-CCSD(T) and a newly developed CBS extrapolation scheme\cite{GriBraBanHan-JCP-2015}, ref. 2 using QCISD(T)/CBS\cite{SedJanPitRezPulHob-ACS-2013} and ref. 3 using LNO-CCSD(T)/CBS(Q,5)) (Local Natural Orbital)\cite{HamNagBarKalBraTka-unknown-2020}. We used ref. 1 of Grimme and co-workers in the main paper. For interaction energies of individual complexes, see Fig. \ref{fig:L7Full}}
\begin{tabular}{l|lll}
MAE         & ref. 1 & ref. 2 & ref. 3 \\ \hline
ref. 1     & 0     & 1.70   & 0.77         \\
ref. 2     & 1.70  & 0      & 1.36      \\
ref. 3     & 0.77  & 1.36   & 0           \\
MP2         & 8.74  & 7.20   & 8.55        \\
SPL         & 3.83  & 2.59   & 3.74        \\
SPL2        & 0.89  & 1.26   & 0.95       \\
MPACF-1     & 2.32  & 1.50   & 2.42       \\
\end{tabular}
\end{table}

\begin{figure}
\includegraphics[width=1\columnwidth]{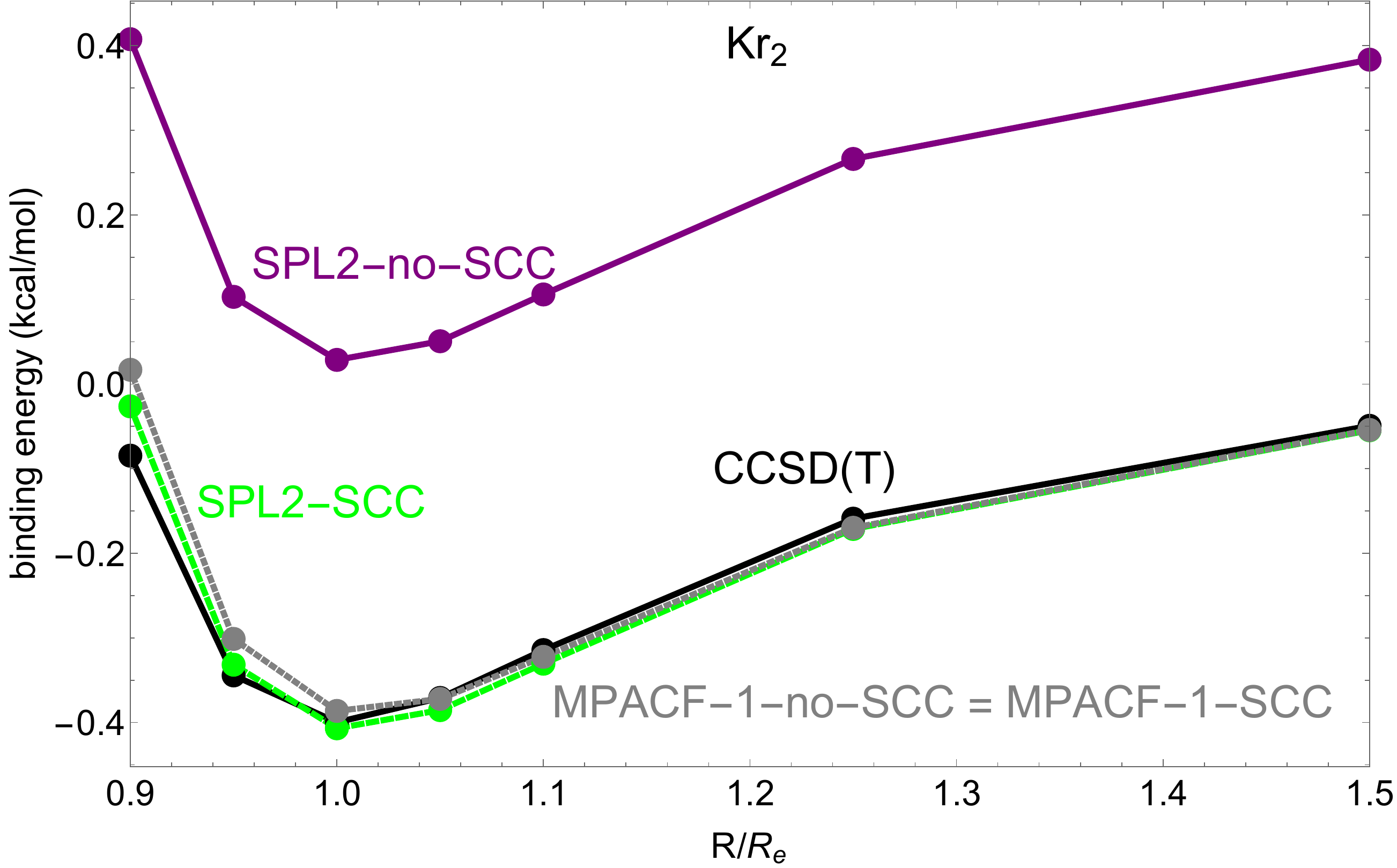}
\caption{The SPL2 with and without size consistency corrections (SCC) plotted for the Kr$_2$ vs the MPACF-1 method with and without SSC and the reference CCSD(T) data. For a complex composed by identical fragments $A$, the following equation, $W^{\rm model}\left (N\textbf{W}(A) \right )=N W^{\rm model}(\textbf{W}(A))$, is a size-extensivity requirement for adiabatic connection model functions, $W^{\rm model}$, with $\textbf{W(A)}=\{W_{1}(A),\ldots,W_{i}(A)$\} being a compact notation for the $i$ input ingredients for fragment $A$ and $N$ the number of fragments\cite{VucGorDelFab-JPCL-18}. SPL2 violates this equation, while MPACF-1 obeys it. Because of that, without the SCC, interaction energies of SPL2 do not vanish even for systems that dissociate into equal fragments as it can be seen from the Kr$_2$ example here. 
Since MPACF-1 is size-extensive, the addition of the SCC does not change the Kr$_2$ dissociation curve, as it is already correct in the dissociation limit. In any case, all our models must be used with the SCC as it ensures that the interaction energies vanish in the dissociation limits (at least for systems dissociating into fragments with non-degenerate ground-state). Without the SCC, meaningless interaction energies would be obtained in some instances [see Ref.~\citenum{VucGorDelFab-JPCL-18}]. 
}
\label{fig:NoSCC}
\end{figure}

\begin{figure}
\includegraphics[width=1\columnwidth]{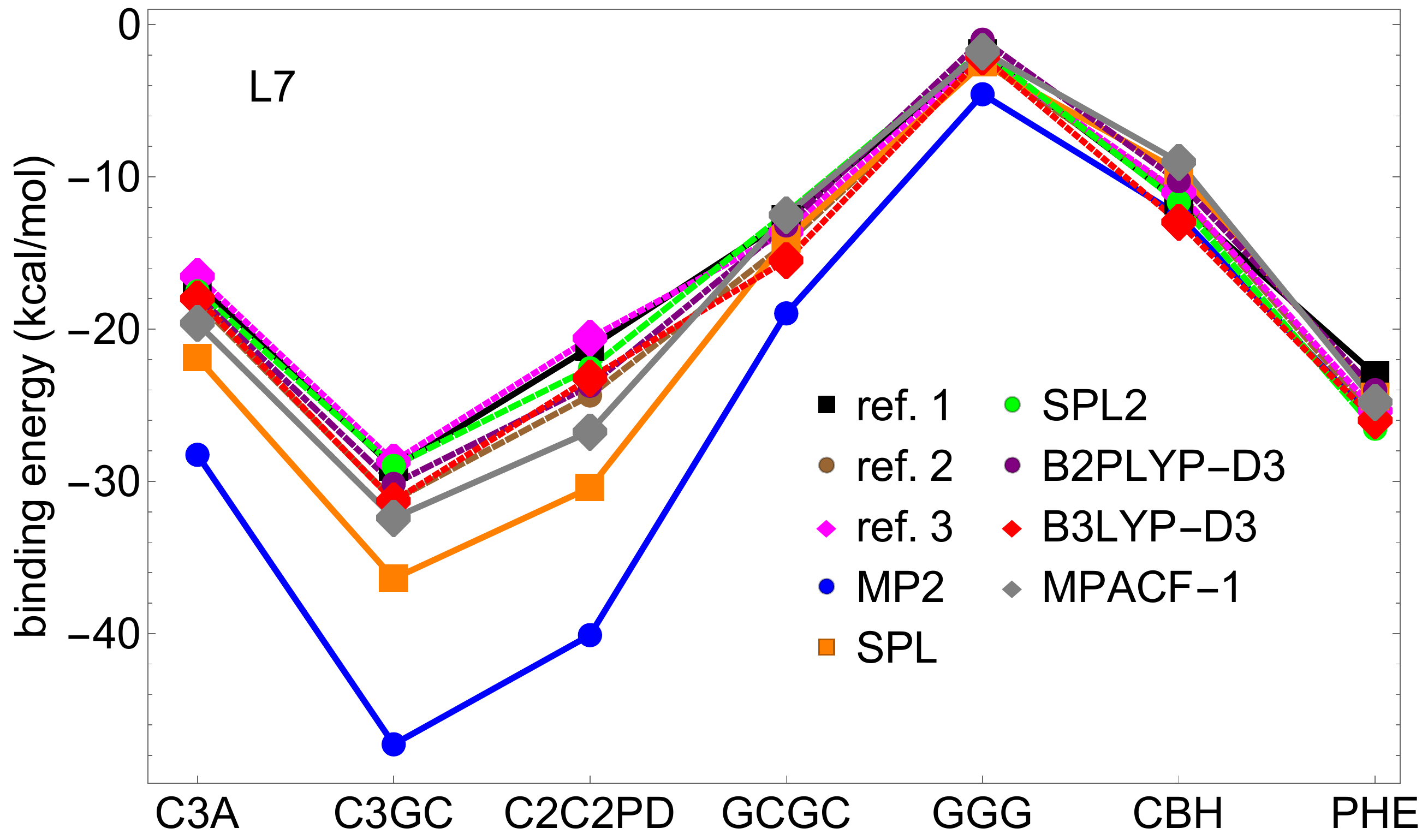}
\caption{ Interaction energies of MP2, SPL, SPL2, B3LYP-D3 and B2PLYP as well as three different reference (ref. 1 is Grimme et al.~\cite{GriBraBanHan-JCP-2015}, ref. 2 is Sedlak et al.~\cite{SedJanPitRezPulHob-ACS-2013} and ref. 3 is Al-Hamdani et al.~\cite{HamNagBarKalBraTka-unknown-2020}) data plotted for all 7 complexes of the L7 dataset. For further details on these references, see Table S1.}
\label{fig:L7Full}
\end{figure}

\begin{figure}
\includegraphics[width=1\columnwidth]{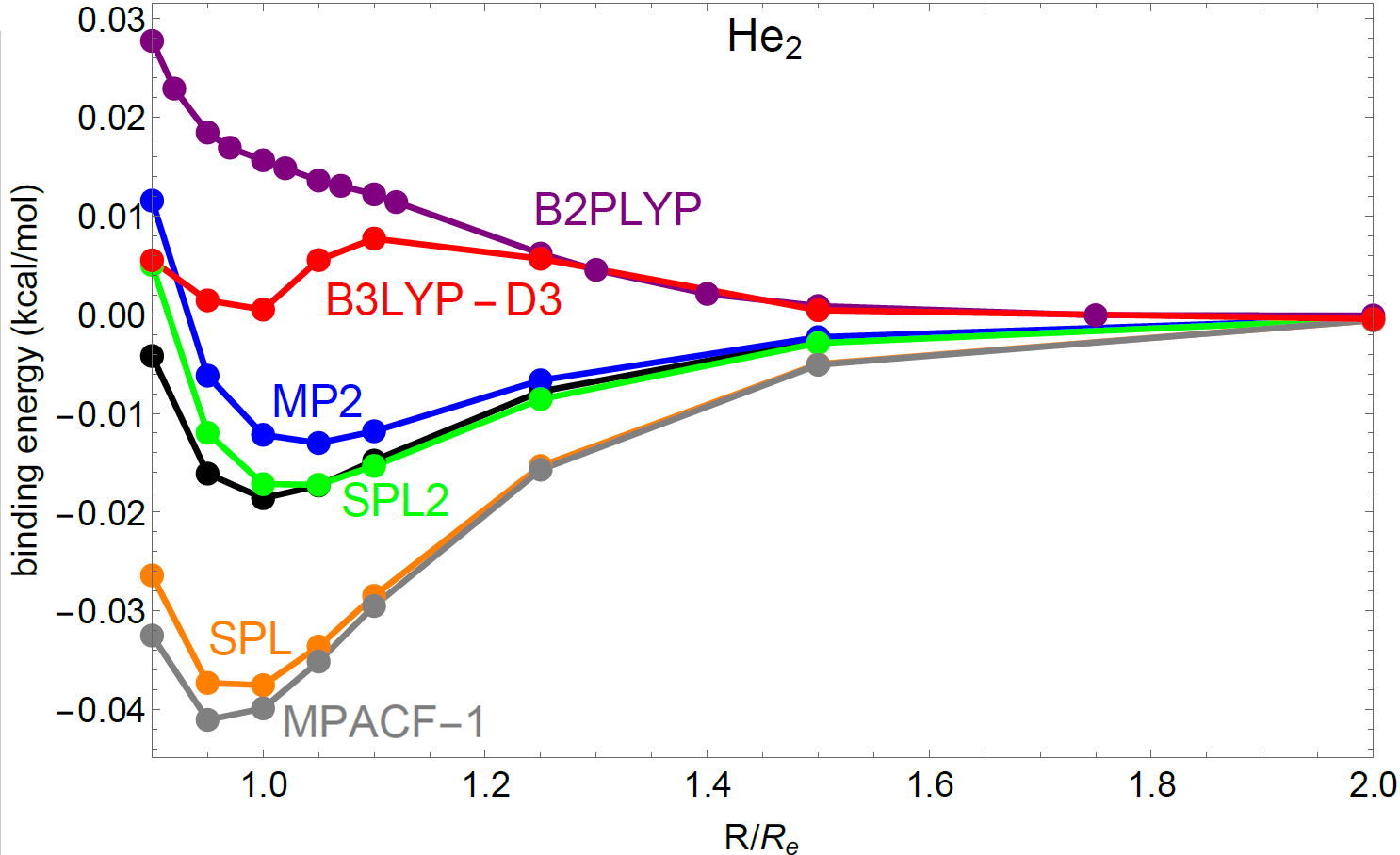}
\caption{The interaction energies of MP2, SPL, SPL2, MPACF-1, B3LYP-D3 and B2PLYP as well as reference CCSD(T) curves for He$_2$. B3LYP even upon addition of D3 is producing an unphysical curve.}
\label{fig:HeFull}
\end{figure}

\begin{figure}
\includegraphics[width=1\columnwidth]{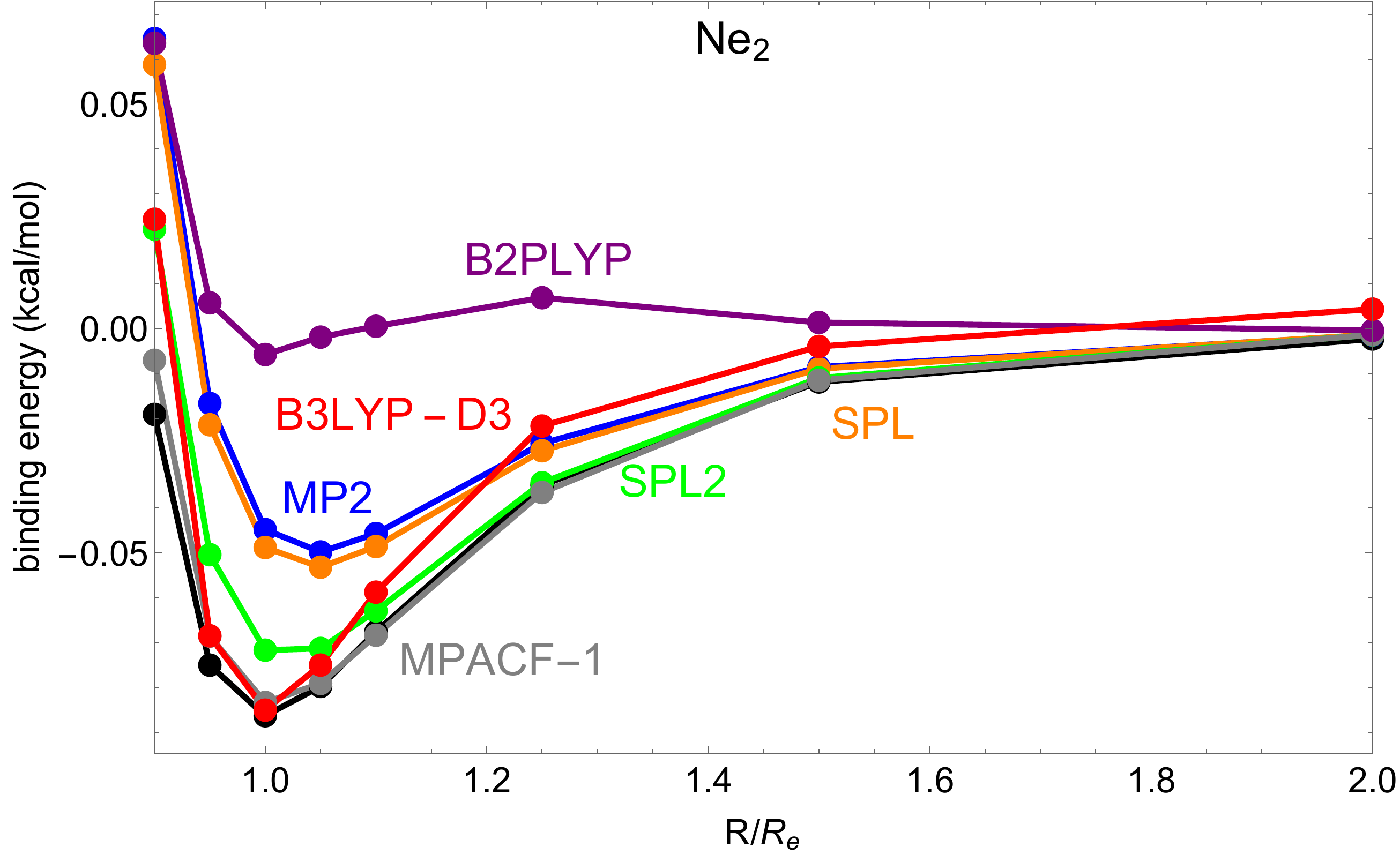}
\caption{The interaction energies of MP2, SPL, SPL2, MPACF-1, B3LYP-D3 and B2PLYP as well as reference CCSD(T) curves for Ne$_2$.}
\label{fig:Ne2}
\end{figure}

\begin{figure}
\includegraphics[width=1\columnwidth]{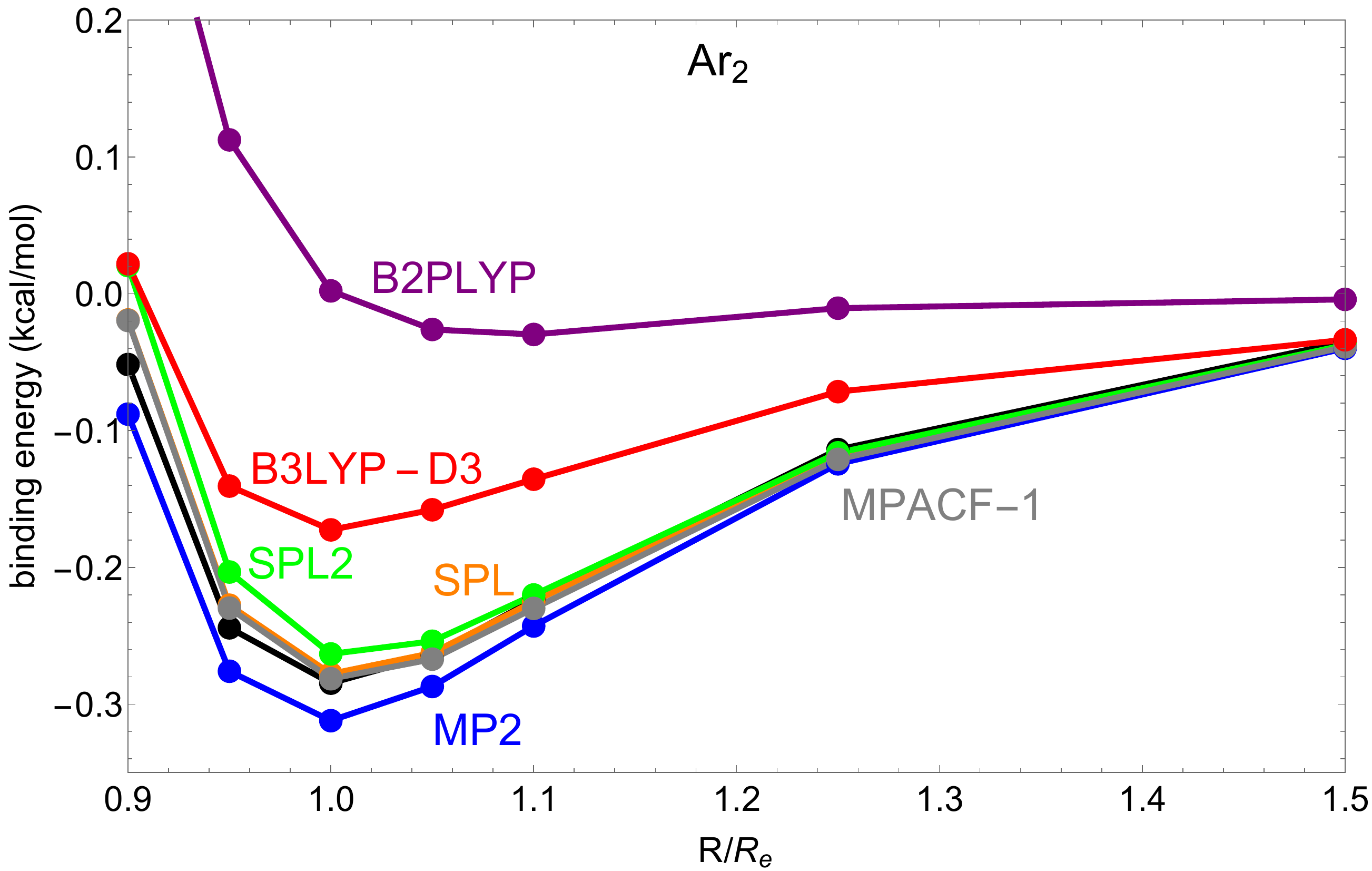}
\caption{The interaction energies of MP2, SPL, SPL2, MPACF-1, B3LYP-D3 and B2PLYP as well as reference CCSD(T) curves for Ar$_2$.}
\label{fig:Ar2}
\end{figure}

\begin{figure}
\includegraphics[width=1\columnwidth]{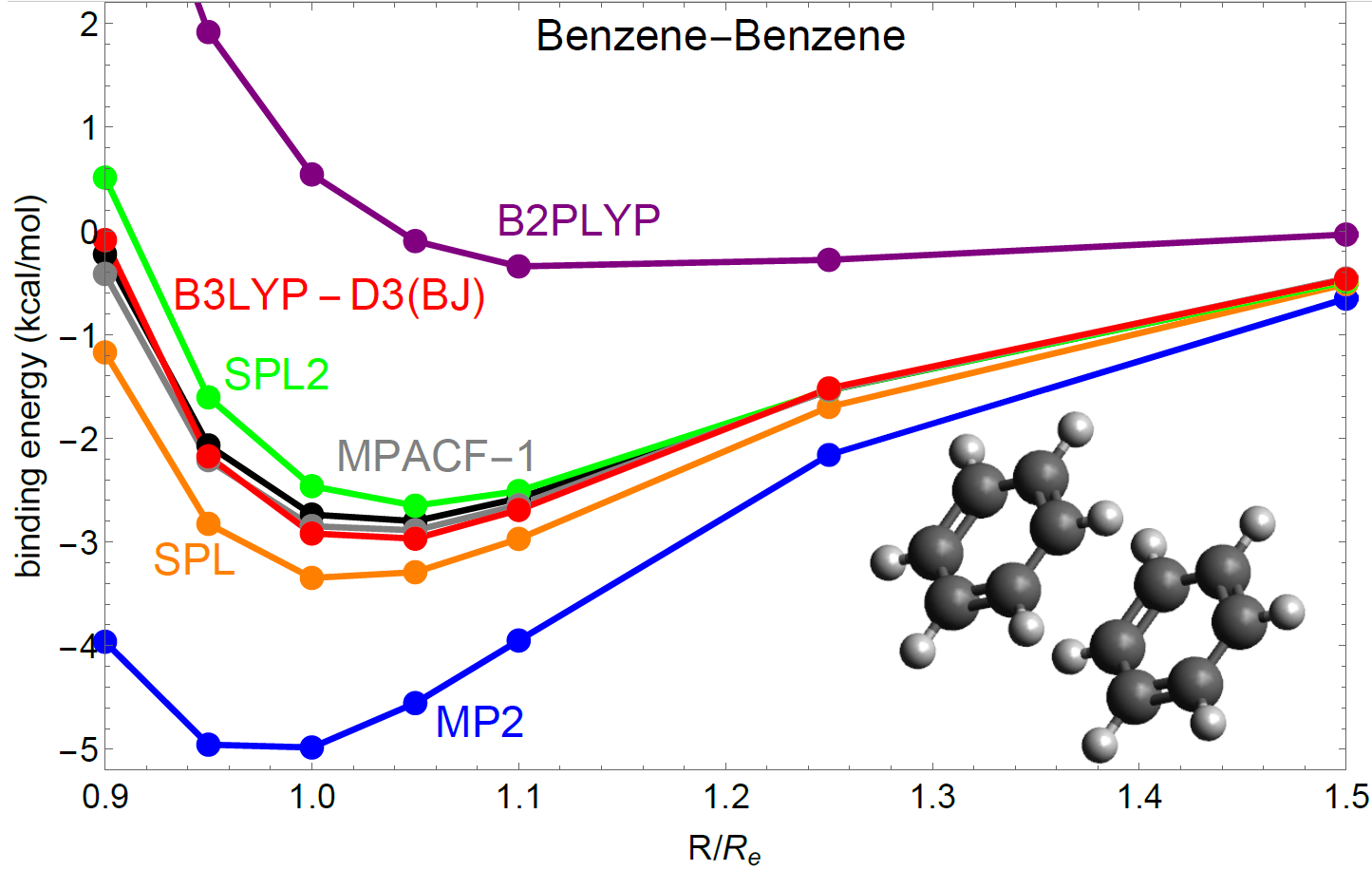}
\caption{The interaction energies of MP2, SPL, SPL2, MPACF-1, B3LYP-D3(BJ) and B2PLYP as well as reference CCSD(T) curves for Benzene dimer.}
\label{fig:Benz}
\end{figure}

\begin{figure}
\includegraphics[width=1\columnwidth]{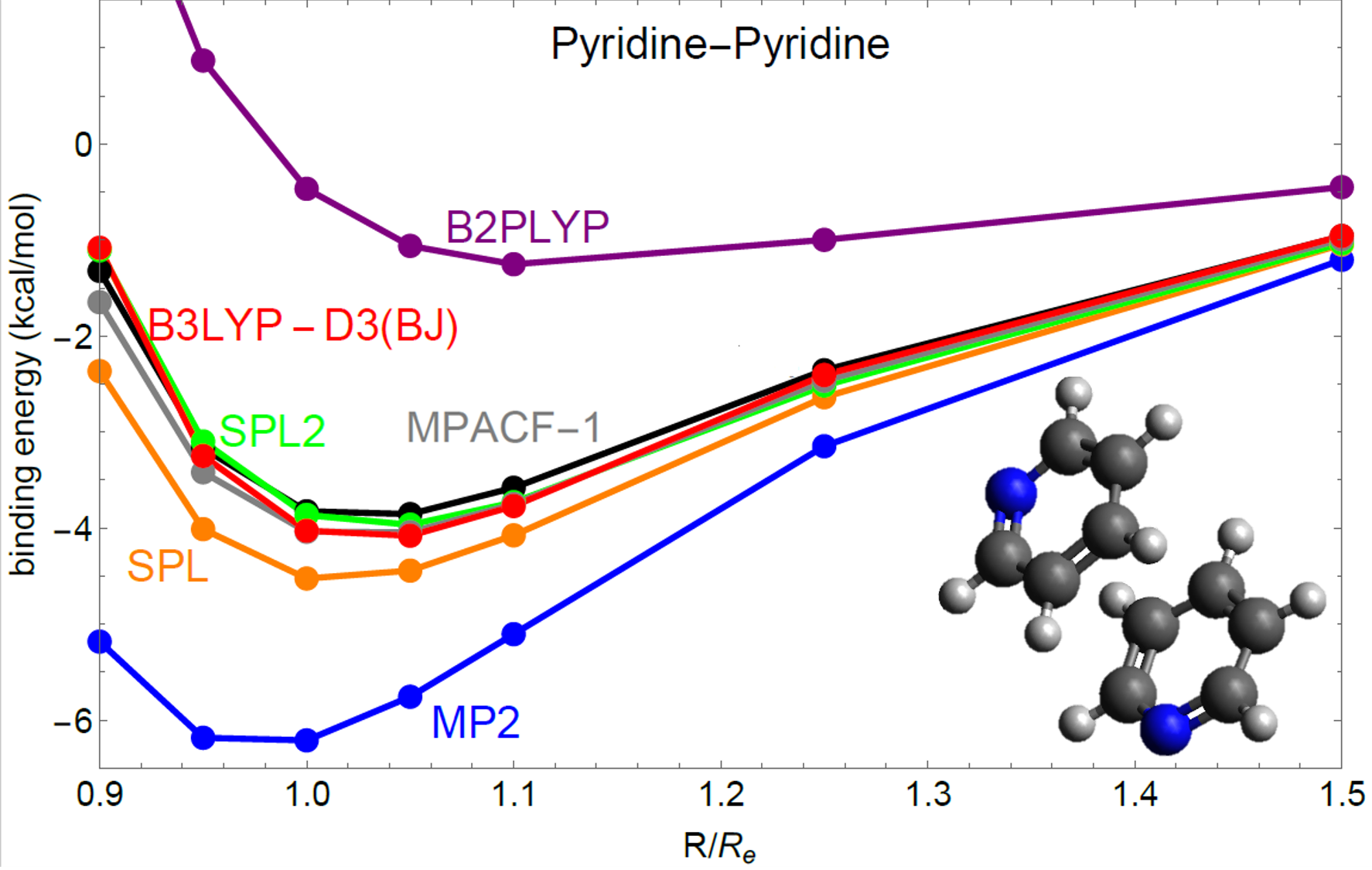}
\caption{The interaction energies of MP2, SPL, SPL2, MPACF-1, B3LYP-D3(BJ) and B2PLYP as well as reference CCSD(T) curves for Pyridine dimer.}
\label{fig:Pyr}
\end{figure}

\begin{figure}
\includegraphics[width=1\columnwidth]{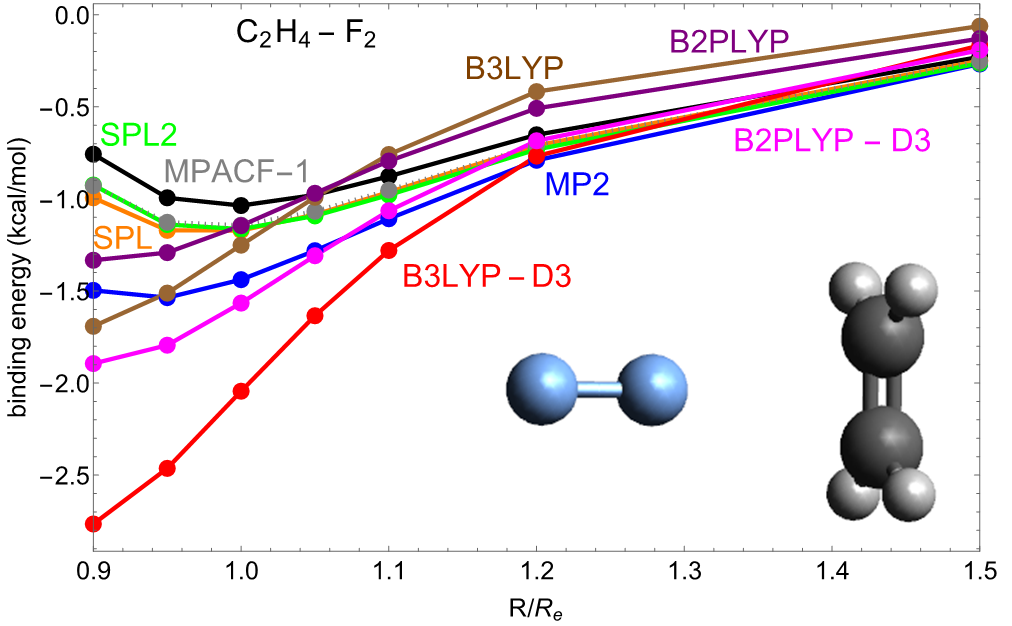}
\caption{The interaction energies of MP2, SPL, SPL2, MPACF-1, B3LYP-D3 and B2PLYP as well as reference CCSD(T) curves for \ce{C2H4}-\ce{F2}}
\label{fig:C2H2}
\end{figure}

\begin{sidewaystable*}[]
\small
\centering
\caption{The $W_{c,\lambda}$ and $E_c$ forms of the three different methods (SPL, SPL2 and MPACF-1) as well as their fixed and empirical parameters. For all forms, $E_x$ and $E^{MP2}_c$ are the exact exchange energy and the MP2 correlation energy respectively, whereas $W^{\rm PC}_{\infty} [\rho^{\rm HF}]$ is a GEA functional ('PC model') evaluated on the HF density, which takes the following integral form: $W^{\rm PC}_{\infty} [\rho^{\rm HF}]=\int \left[A\rho^{\rm HF}(\mathbf{r})^{4/3} + B\frac{|\nabla \rho^{\rm HF}(\mathbf{r})|^2}{\rho^{\rm HF}(\mathbf{r})^{4/3}}\right]\mathrm{d}\mathbf{r}$ with $A=-1.451$ and $B=5.317\times10^{-3}$. In SPL, $W_{c,\infty}$ form results from the PC model approximation to $W_{c,\infty}^{\rm DFT}$ as done in Ref.~\citenum{VucGorDelFab-JPCL-18}. In SPL2, $W_{c,\infty}$ has been approximated in terms of the form containing $\alpha$ and $\beta$ parameters, which have been determined empirically. The form of $W_{c,\infty}$ in MPACF-1 has been fixed by the large $\lambda$ limit of the MP AC for the uniform electron gas.}
\begin{tabular}{|c|c|c|c|}
\hline
& SPL & SPL2 & MPACF-1\\ \hline
 $W_{c,\lambda}$ & $W_{c,\infty} 
\left (1 -  \frac{1}{\sqrt{1 + b \lambda} }\right)$ & $ C_1 -  \frac{m_1}{\sqrt{1 + b_1 \lambda} } - \frac{m_2}{\sqrt{1 + b_2 \lambda} }$ & $g \left(\frac{(h+1) \left(h \sqrt{d_1^2 \lambda+1} \left(3 d_2^4 \lambda+4\right)+2 \left(d_1^2 \lambda+2\right) \left(d_2^4 \lambda+1\right)^{3/4}\right)}{4 \sqrt{d_1^2 \lambda+1} \left(d_2^4
   \lambda+1\right)^{3/4} \left(\sqrt{d_1^2 \lambda+1}+h \sqrt[4]{d_2^4 \lambda+1}\right)^2}-1\right)$ \\ \hline
 $E_{c}=\int_{0}^{1} W_{c,\lambda} d\lambda$& $W_{c,\infty}  \left(\frac{2 + b - 2 \sqrt{1 + b}}{b}\right)$ & $C_1 -  \frac{2m_1\left(\sqrt{1 + b_1}-1\right)}{b_1}- \frac{2m_2\left(\sqrt{1 + b_2}-1\right)}{b_2}$  & $-g +\frac{g (h+1) }{\sqrt{d_1^2  +1}+h \sqrt[4]{d_2^4 +1}}$ \\ \hline
 $W_{c,\infty}$ & $W_{\infty}^{\rm PC}[\rho ^ {\rm HF}] - E_{x}$& $\alpha W^{\rm PC}_{\infty} [\rho^{\rm HF}] + \beta E_{x}$ & $\alpha W^{\rm PC}_{\infty} [\rho^{\rm HF}] + \beta E_{x}$ \\ \hline
  \text{Fixed params.} & $b=\frac{4 E_{\rm c}^{\rm MP2}}{W_{c,\infty}}$ & $C_1=W_{c,\infty}$, $b_1 = \frac{b_2~m_2- 4 E_c^{\rm MP2}}{m_2 - W_{c,\infty}}$, $m_1 = W_{c,\infty} - m_2$& $g=-W_{c,\infty}$, $ h=\frac{\displaystyle 4E_{c}^{\rm MP2} - 2 d_1^2 W_{c,\infty}}{\displaystyle -4 E_{c}^{\rm MP2} + d_2^4 W_{c,\infty}}$ \\ \hline
 \text{Emp. params.} & - & $b_2=0.117$, $m_2=10.68$, $\alpha=1.1472$, $\beta=-0.7397$ & $d_1=0.294$, $d_2=0.934$ $\alpha=1$, $\beta=1$ \\ \hline
\end{tabular}
\end{sidewaystable*}

 \newpage

\bibliography{bib_clean,sh8}